# Survey techniques, detection probabilities, and the relative abundance of the carnivore guild on the Apostle Islands (2014-2016)

## Final Report
December 15, 2016

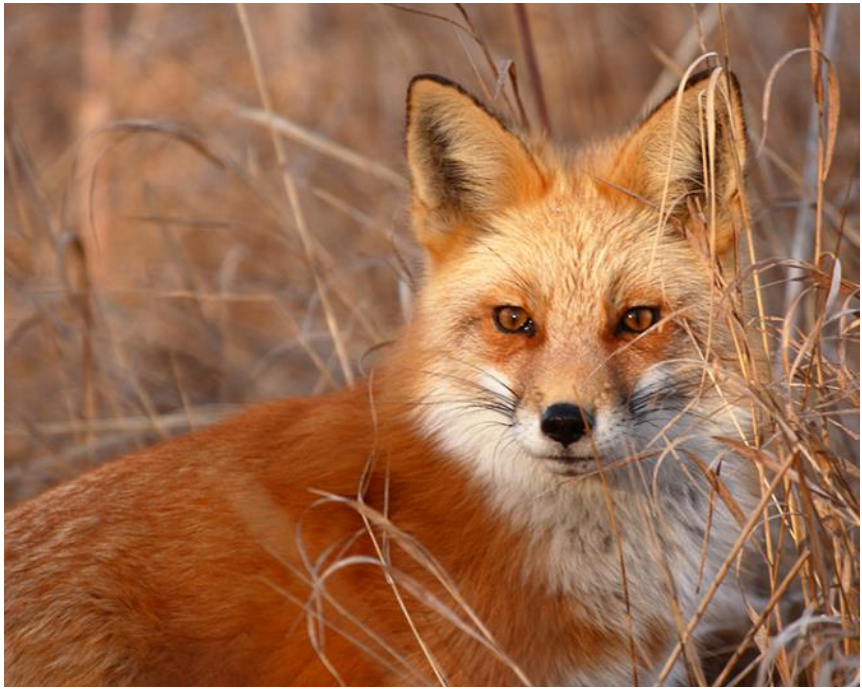

Maximilian L. Allen – *Department of Forest & Wildlife Ecology, University of Wisconsin, Madison*
Bryn E. Evans – *Department of Forest & Wildlife Ecology, University of Wisconsin, Madison*
Michael E. Wheeler – *Department of Forest & Wildlife Ecology, University of Wisconsin, Madison*
Marcus A. Mueller – *Department of Forest & Wildlife Ecology, University of Wisconsin, Madison*
Kenneth Pemble – *Resource Management and Planning, Apostle Islands National Lakeshore*
Erik R. Olson – *Biology and Natural Resources, Northland College, Ashland Wisconsin*
Julie Van Stappen *(Principal Investigator)* – *Resource Management and Planning, Apostle Islands National Lakeshore*
Timothy R. Van Deelen *(Principal Investigator)* – *Department of Forest & Wildlife Ecology, University of Wisconsin, Madison*

Cover photograph by Maximilian Allen

## SUMMARY AND AKNOWLEDGEMENTS

This is the final report for Great Lakes Northern Forest CESU Cooperative Agreement P14AC01180 between the United States Department of Interior and University of Wisconsin, Madison, entitled "Survey Carnivores at Apostle Islands National Lakeshore using Remote Cameras."

The Apostle Islands carnivore guild assessment project was a 3-year cooperative investigation between researchers at the University of Wisconsin at Madison, Northland College, and the National Park Service. Researchers from all three groups contributed to the study design, along with the placement and maintenance of cameras. Researchers at the University of Wisconsin provided statistical analyses and the write-up, and will ensure that results will be published in peer-reviewed journals.

We give many thanks to the personnel from each group that contributed to this project, especially the numerous technicians from the National Park Service, graduate students from the Van Deelen lab at the University of Wisconsin-Madison, and students from Northland College that provided assistance over the course of the project. The support and cooperative spirit between the three groups was a key aspect of this project being accomplished.

Suggested Citation:
Allen, M. L., B. E. Evans, M. E. Wheeler, M. A. Mueller, K. Pemble, E. R. Olson, J. Van Stappen, and T. R. Van Deelen. 2016. Survey techniques, detection probabilities, and the relative abundance of the carnivore guild on the Apostle Islands (2014-2016). Final Report to the National Park Service.

# ABSTRACT

Carnivores are important components of ecosystems with wide-ranging effects on ecological communities. Specific effects of carnivore species on ecological communities are complex and vary with size, natural history, and hunting tactics, and both researchers and managers must understand ecological roles of carnivores and how they interact with their local environment. We studied the carnivore community in the Apostle Islands National Lakeshore (APIS), where the presence, distribution, and populations of carnivores was largely unknown and were needed to better understand the island-level variation in presence, distribution, and composition of carnivore communities, and how this affected the ecology of APIS. We developed a systematic method to deploy camera traps across a grid while targeting fine-scale features to maximize carnivore detection (Appendix 1), including systematic methods for organizing and tagging the photo data (Appendix 2).

In this report we document our findings from deploying 88 cameras on 13 islands from 2014-2016. We collected 92,694 photographs across 18,721 trap nights, including 3,591 wildlife events and 1,070 carnivore events. We had a mean of 6.6 cameras per island (range 2-30), and our camera density averaged 1.23 (range 0.74-3.08) cameras/ km$^2$. We detected 27 species and 10 terrestrial carnivores, including surprising detections of American martens (*Martes americana*) and gray wolves (*Canis lupus*). Our observations of American martens, the only state endangered mammal in Wisconsin, are of particular interest because their presence was uncertain and their discovery may inform recovery efforts on the mainland. The mean richness of carnivores on an island was 3.23 (range 0-10). The best single variable to explain carnivore richness on the Apostle Islands was island size, while the best model was island size (positive correlation) and distance from mainland (negative correlation) ($R^2 = 0.92$). Relative abundances for carnivores ranged from a low of 0.01 for weasels (*Mustela spp.*) to a high of 2.64 for black bears (*Ursus americanus*), and the relative abundance of a species was significantly correlated with the number of islands on which they were found. Carnivore occupancy ranged from lows of 0.09 for gray wolves and 0.11 for weasels to a high of 0.82 for black bears. Detection rates were significantly higher in summer than winter, ranging from 0.38 for red fox to 0.88 for black bear in summer, and from 0.01 for weasel to 0.20 for coyote in winter.

Low levels of human development and recreation in APIS may play a role in supporting carnivore species that avoid human disturbance. However, none of the islands in the archipelago are likely large enough to sustain populations of mammalian carnivores in the face of demographic stochasticity or the genetic effects of small population size. Hence, one important area for future study is determining how carnivores colonize and move between islands, as well as how the carnivore community interacts and effects each other. Fuller understanding of APIS ecology will require on-going monitoring of carnivores to evaluate temporal dynamics as well as related ecological evaluations (e.g. small mammal dynamics, plant community dynamics) to understand trophic effects.

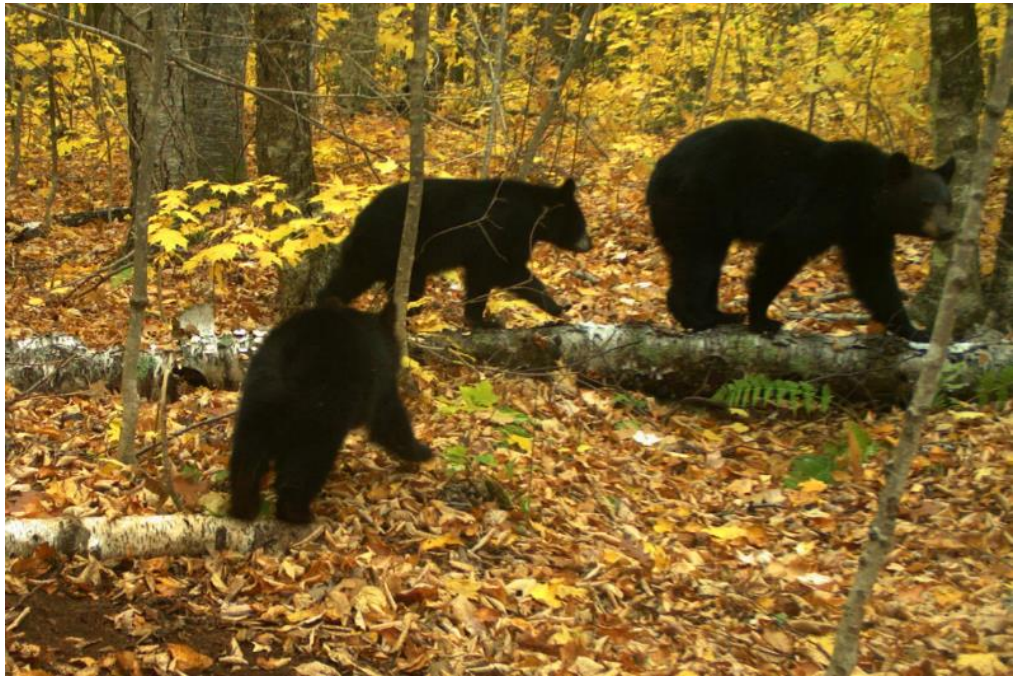

*Photo 1. Black bear cubs with their mother during autumn on Hermit Island.*

# INTRODUCTION

Carnivores are important components of ecosystems with wide-ranging effects on ecological communities. Carnivores affect composition and structure of ecological communities (Estes and Palmisano 1974, McLaren and Peterson 1994, Estes 1996, Allen et al. 2014). Effects of carnivores on prey occur both directly (i.e., through predation; Estes 1996, Ripple et al. 2014) and indirectly (i.e., causing shifts in habitat selection to avoid predation; Brown et al. 1999, Altendorf et al. 2001, Ripple and Beschta 2004, Atwood et al. 2007). In addition, carnivores may affect composition and abundance of non-prey species in communities (Estes and Palmisano 1974, Hunter and Price 1992, Courchamp et al. 1999, Prugh et al. 2009, Allen et al. 2015).

Given their effects on ecological communities, loss of carnivores potentially may change ecosystem dynamics. However, worldwide and throughout the USA, many carnivore populations are threatened and diminished (Laliberte and Ripple 2004, Ripple et al. 2014). Carnivores are among the most charismatic wildlife species (Kellert 1997) and are important to consider in the management of National Parks. The National Park System (NPS) mission is to "*preserve unimpaired the natural and cultural resources and values of the NPS for the enjoyment, education, and inspiration of this and future generations*" (Anderson and Barbour 2003).

Despite apparent similarities in ecological roles of carnivores, different species occupy different ecological niches, and their influences depend on ecological context. Specific effects of carnivore species on ecological communities vary with size, natural history, and hunting tactics (Ripple et al. 2014), but also sometimes depend on interactions with each other. For example, carnivores will preferentially associate with areas inhabited by prey species while simultaneously avoiding habitats occupied by sympatric but dominant competitors (Lesmeister et al. 2015, Wang et al. 2015). Intraguild competition can lead to trophic cascades, wherein apex carnivores release subordinate small carnivores from competitive pressure by reducing the abundance of dominant mesocarnivores (Allen et al. 2015, Lesmeister et al. 2015, Wang et al. 2015). Effects of carnivores on ecosystems are therefore complex and varied, and both researchers and managers must understand ecological roles of carnivores and how they interact with their local environment.

Monitoring distribution and trends in population sizes of species are a fundamental part of wildlife management, both in protected (i.e., parks) and non-protected areas, especially for charismatic and ecologically important species like mammalian carnivores. Carnivores, however, can be difficult to monitor due to their low population densities and cryptic behaviors (Harmsen et al. 2010, Krofel et al. 2012, Allen et al. 2016), making rigorous monitoring difficult. This difficulty is exacerbated when, as in the Apostle Islands National Lakeshore (APIS), areas for survey are remote and difficult to access when weather makes travel dangerous.

APIS was established in 1970 and included 21 of the 22 islands to protect their unique cultural and natural value (Busch 2008). Human use is limited to recreational and land management activities (Feldman 2004), but the presence, distribution, and populations of carnivores on the APIS was largely unknown. Recent wildlife research included black bear (*Ursus americanus*) population dynamics (Belant et al. 2005), which found substantial black bear immigration from mainland populations. A historical observational report recorded the presence of red fox (*Vulpes vulpes*) on the islands, and coyotes (*Canis latrans*) were observed traveling on the ice (Jackson 1920). Reintroductions of American martens (*Martes americana*) occurred in APIS during the 1950's (Williams et al. 2007), but these reintroductions are assumed to be a failure (Williams et al. 2007).

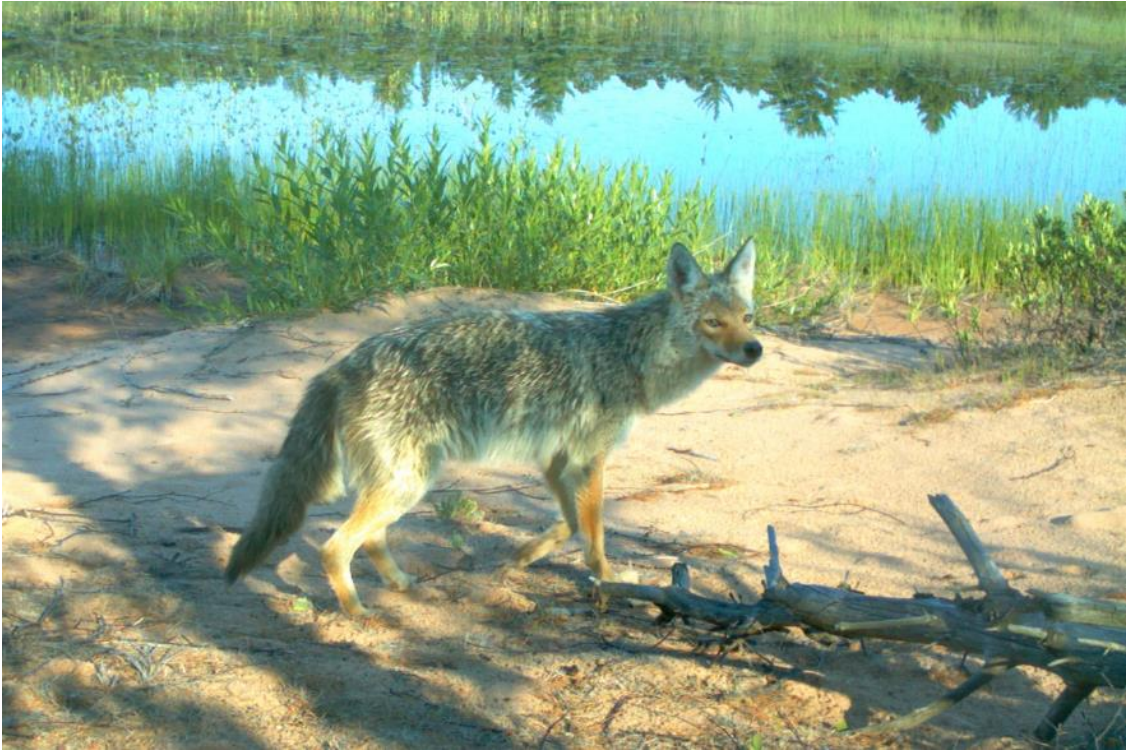

*Photo 2. A coyote walking along a lagoon on Stockton Island.*

Formal studies of carnivore distribution in the archipelago were needed to better understand the island-level variation in presence, distribution, and composition of carnivore communities, and how this affected the ecology of APIS. Important effects on community ecology include how carnivores may affect the white-tailed deer (*Odocoileus virginianus*) population. For example, Canada yew (*Taxus canadensis*) is a plant that is increasingly diminished in Wisconsin primarily due to overbrowsing by deer, and carnivore regulation of deer density, may facilitate the persistence of relict yew populations in APIS. It also is important to understand which carnivore species are present in APIS. For example, American martens are currently the only state endangered mammal in Wisconsin (Woodford and Dunyham 2011), and despite anecdotal reports, it is unknown if they exist in APIS.

Surveys performed via camera-traps (i.e., motion-triggered cameras) are a potential solution to the difficulty of monitoring the carnivore community in APIS. The optimal survey design and analytical techniques for camera trapping vary widely and efforts to develop and standardize optimal procedures are ongoing (see review in Burton et al. 2015). The design used depends critically on the monitoring goals and the logistical constraints associated with camera deployment and data recovery. This project builds on a previous cooperative agreement between the USDI NPS and the University of Wisconsin-Madison to design a rigorous monitoring program for large carnivores in APIS. Our original project objectives, as noted in the grant proposal and grant agreement between the University of Wisconsin-Madison and USDI NPS were:

    1) Develop protocols for using trail cameras to determine the occurrence of carnivores, including relative abundance, where possible. Ensure that protocols can be used throughout the park and are readily transferrable to other remote areas.

    2) As part of protocol development, obtain: estimates of detection probabilities for each carnivore species to understand how long cameras need to be deployed; determine the season of greatest detection probability to understand time of year cameras should be deployed; and through sub-sampling of cameras, know the camera density needed to detect species of interest.

    3) For the pilot location, Stockton Island, determine which carnivore species are present and their distribution. If possible, determine relative abundance.

In consultation with APIS staff, we have expanded beyond the scope of these objectives, including expanding our monitoring efforts beyond Stockton Island.

Additional goals of the project were to examine the dynamics of the islands' carnivore guild using occupancy modeling and infer how these processes fit into the context of island biogeography and community ecology. In this report we document our findings from deploying 88 cameras on 13 islands from 2014-2016 (Figure 1). This includes a) the species richness of each of the islands we monitored, b) the distribution and relative abundance of carnivore species, c) the occupancy and detection rates of each carnivore species. We also provide methodology we developed for setting up a camera monitoring project (Appendix 1), including systematic methods for organizing and tagging the photo data (Appendix 2). We conclude with recommendations on methods and areas of research for the future.

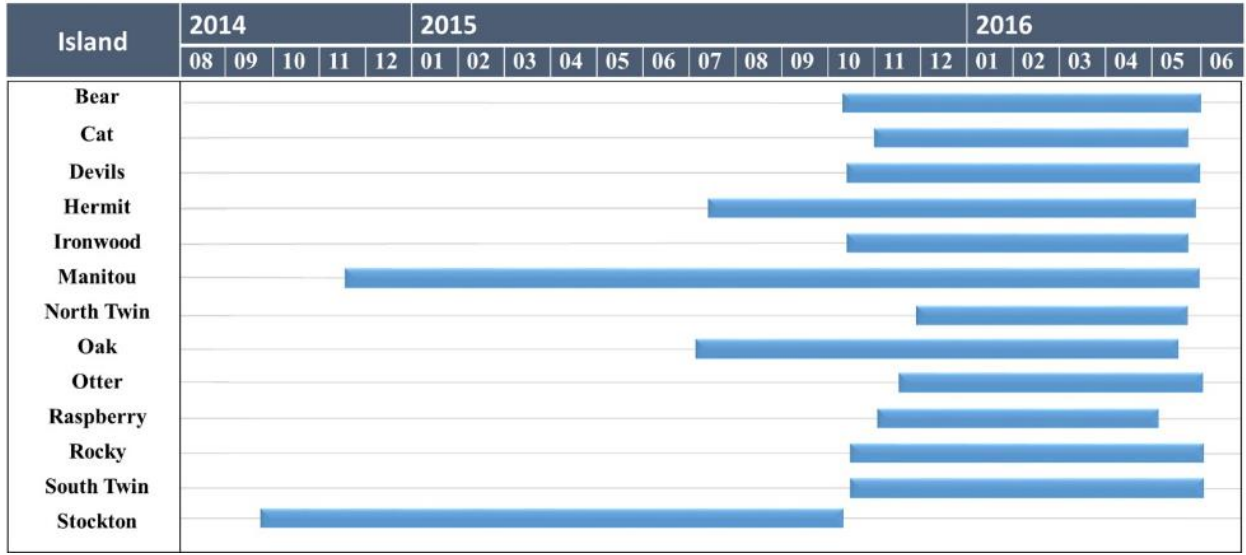

Figure 1. Dates of camera deployment in the Apostle Islands, Wisconsin (USA).

## METHODS

### Study Area

The Apostle Islands are an archipelago of 22 of primarily Precambrian sandstone islands located in southwest Lake Superior off the northern tip of the Bayfield Peninsula in southwestern Lake Superior, Wisconsin, USA. The islands are in a transitional zone between northern boreal coniferous forest and deciduous forest (Craven and Lev 1987). Mean maximum and minimum monthly temperatures on the islands were 24.7°C and -14.4°C in July and January respectively, and annual precipitation was approximately 75 cm, with approximately 200 cm of snow (National

Climatic Data Center 2011, Thayn 2013). Our study occurred on 13 islands (see Table 1 for geographic characteristics and sampling effort).

*Table 1. Characteristics of the individual islands and sampling effort for remotely triggered cameras deployed within the Apostle Islands, Wisconsin (USA, 2014-2016).*

| Island | Cameras Deployed | Total Trap Nights | Island Size (km2) | Distance to Mainland (km) | Distance to Nearest Island (km) | Maximum Elevation (m) | Mean Elevation (m) |
|---|---|---|---|---|---|---|---|
| Bear | 8 | 1392 | 7.34 | 7.23 | 2.84 | 72 | 26.90 |
| Cat | 5 | 766 | 5.41 | 18.03 | 2.74 | 25 | 13.3 |
| Devils | 2 | 464 | 1.25 | 14.33 | 3.36 | 21 | 10.63 |
| Hermit | 3 | 685 | 3.17 | 3.67 | 2.20 | 56 | 21.7 |
| Ironwood | 2 | 277 | 2.69 | 14.44 | 1.66 | 27 | 15.26 |
| Manitou | 4 | 647 | 5.36 | 8.43 | 1.66 | 43 | 19.7 |
| North Twin | 2 | 356 | 0.65 | 20.76 | 2.73 | 13 | 8.44 |
| Oak | 16 | 4585 | 20.32 | 2.12 | 2.22 | 147 | 66.8 |
| Otter | 5 | 1009 | 5.35 | 8.43 | 1.29 | 44 | 24.4 |
| Raspberry | 2 | 365 | 1.16 | 2.69 | 2.91 | 30 | 15.4 |
| Rocky | 5 | 816 | 4.24 | 12.41 | 1.05 | 31 | 14.42 |
| South Twin | 2 | 101 | 1.36 | 15.06 | 1.05 | 15 | 8.33 |
| Stockton | 30 | 7258 | 40.00 | 7.84 | 2.15 | 61 | 25.7 |

*Field Methods and Design*

Careful and deliberate camera placement is critical for efficiently documenting carnivores with camera traps (Harmsen et al. 2010, Krofel et al. 2012, Allen et al. 2016) and estimating their distribution and abundance (Chandler and Royle 2013, Burton et al. 2015). We developed a systematic method to deploy camera traps across a 1x1 km grid, while explicitly targeting fine-scale features to maximize carnivore detection (i.e. camera height, orientation, and distance to game trail). We developed a written protocol for use during the study and for future studies (monitoring or research) in APIS and the rest of the NPS (Appendices 1 & 2). We conducted camera trapping year-round to encompass changes in carnivore activity and visual obstructions caused by changes in vegetation.

To summarize our camera deployment methods, we used our protocol to place 88 HC600 Hyperfire™ High Output Covert infrared digital game camera (RECONYX, Inc., Holmen, WI, USA) on 13 islands (Table 1). We programmed cameras to take a photo when triggered by an animal and also to take a "time-lapse" photo every day at 11 am to create a systematic sampling of changing ecological conditions (see below). These cameras had an infrared flash range of 15 m, trigger speed of 1/5 sec, 1080p high definition image resolution. For each photo we programmed the cameras to record the time, date, temperature (°F), and moon phase. We initially programmed the cameras on Stockton Island to take 5 photos, 1 sec between each photo, and a 15 sec delay between events. We then changed our programming to take 3 photos, with 1 min refractory period between each event as we expanded our camera grid to other islands. We marked and recorded the coordinates of each camera site with a handheld GPS unit, but did not place flagging or physically mark any of the camera sites. We returned to each of the camera sites approximately six months after the initial deployment date to replace batteries and memory cards.

For Stockton Island, 34 grid points were generated. Only 30 cameras were available, so we randomly omitted four deployment locations from the final configuration. After additional cameras became available, the restrictions of the 1 x 1 km grid were relaxed to increase the camera density and provide more in-depth coverage of the smaller islands (Figure 2).

We developed a photo-tagging protocol for use in interpreting and analyzing data obtained from photo-traps. This protocol was used during the study and can also be used for future studies in APIS and the rest of the NPS. Our protocol, using MapView Professional (RECONYX, Inc., Holmen, WI, USA) is explained in detail in Appendix 2.

We defined a carnivore photo event as any series of 3-5 photos (as programmed) triggered by a carnivore species. We used the carnivore photo events to determine the relative abundance of carnivores and their presence at each camera. To reduce pseudo-replication when calculating relative abundance, we considered multiple photos of a species within 30 min of a previous photo to be the same event (Naing et al. 2015, Wang et al. 2015).

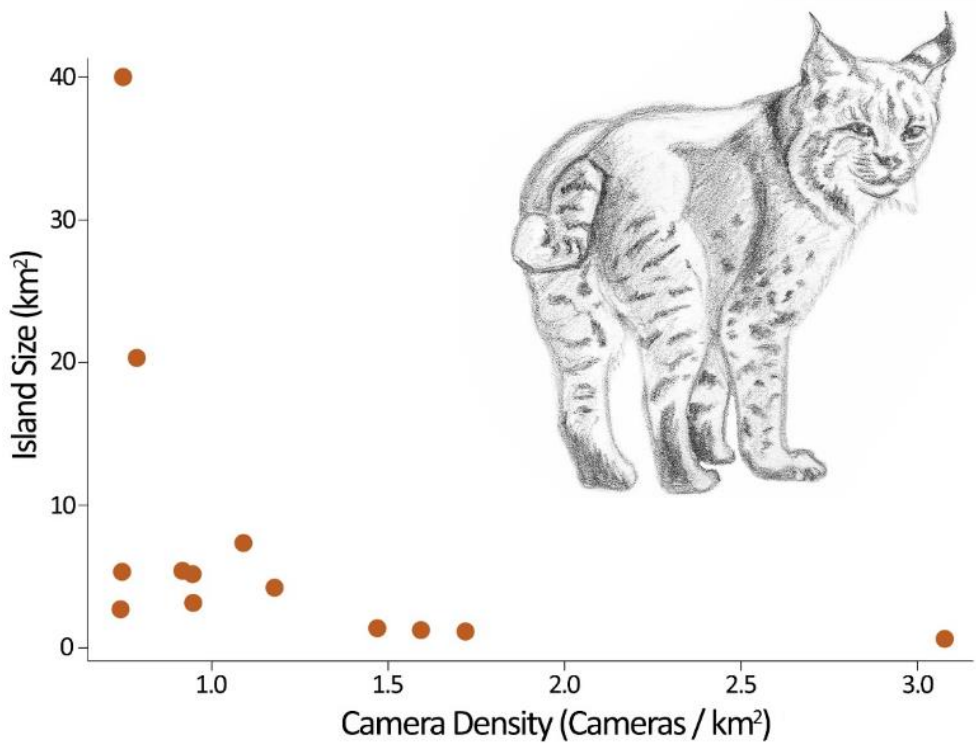

*Figure 2. The relationship of our camera density and island size (drawing by Yiwei Wang).*

***Statistical Analyses***

We used program R version 3.3.1 (R Core Team 2016) for all of our statistical analyses, and in each analysis we considered $p < 0.05$ to be statistically significant.

We calculated carnivore species richness as the number of carnivore species we detected on a given island. We then tested for patterns in carnivore species richness using linear regression. We used the carnivore species richness as our dependent variable and used island size ($km^2$), distance to mainland (km), distance to nearest island (km), maximum elevation (km), and the relative abundance of people as our independent variables.

We calculated relative abundance (*RA*) of a carnivore species as

$$RA = (D / TN) \times 100$$

where *D* is the number of detections and *TN* is the total number of trap nights. We then tested whether the mean relative abundance of a carnivore was correlated with the number of islands the carnivore was found on using a linear regression. Last, we determined the distribution of each terrestrial carnivore species across the islands based on the distributions of cameras with a positive detection.

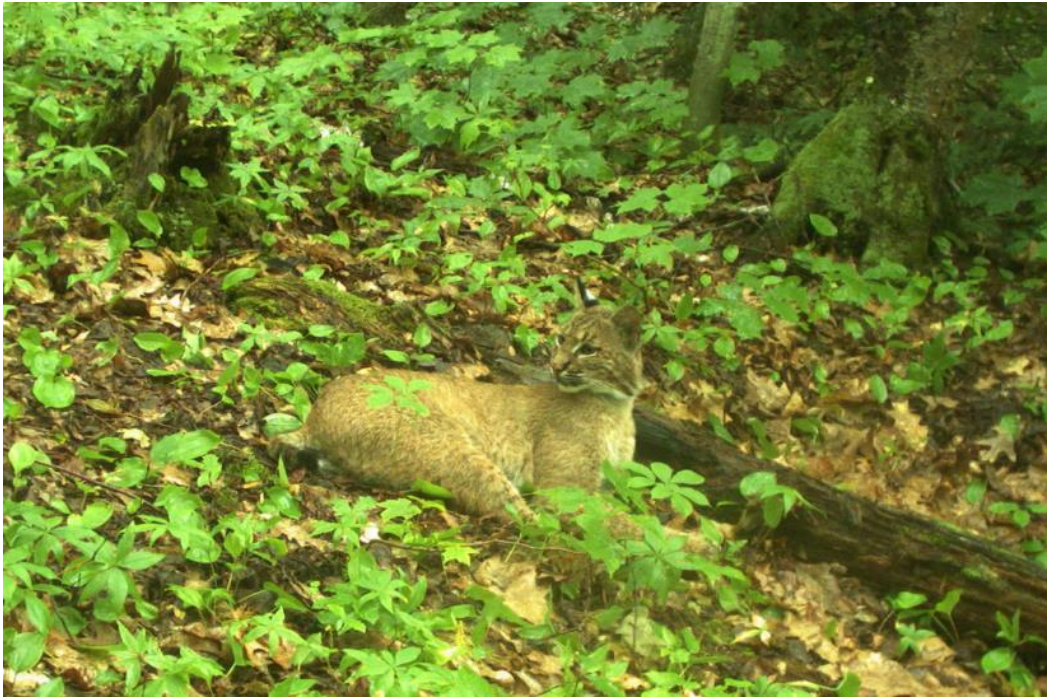

*Photo 3. A bobcat resting in a forest on Oak Island.*

We used occupancy modeling to assess overall status of carnivore species across our multiple camera stations. Occupancy modeling leverages the fact that each observation results from 1) the species being present (*z*) and 2) it being successfully detected. Positively identified images of each species from each camera yielded a site- and species-specific string of zeros (undetected) and ones (detected), which is referred to as a capture history. These capture histories are the inputs to calculate detection probabilities (*p*) and occupancy estimates (Ψ) (MacKenzie et al. 2002, 2006, Royle and Nichols 2003). If a species is not observed, this can result from either its absence ($z_i = 0$) or a non-detection ($z_i = 1$ but $p_i = 0$). Explicitly modeling the probability of a true-absence versus false-absence reduces the bias that can result from assuming perfect detection (White 2005) to calculate occupancy estimates (MacKenzie et al. 2002, 2006, Royle and Nichols 2003).

We consolidated image data as either detected (1) or not detected (0) for each carnivore species during two-week observation periods. Cameras needed to be operating each day of the two-week period to be included in the analysis. This provided discrete detection histories to use for repeated survey methods, while also avoiding autocorrelation between detection only a day apart, and issues associated with zero-inflation. We subdivided the monitoring periods into two seasons: summer (May through October) and winter (November through April). After calculating occupancy and detection rates, we used R package *unmarked* (Fiske and Chandler 2011) and the *ranef* command to calculate occupancy estimates for each camera site. This yielded occupancy estimates where Ψ = 1 for sites with confirmed species presence, and 0 < Ψ < 1 for any site where a species was present but not detected. These estimates can be used to assess differences in species occupancy between islands. The occupancy model for raccoons (*Procyon lotor*) did not converge because of small sample size of observations, and we therefore excluded their occupancy results and subsequent analyses.

We then performed subsequent analyses of occupancy and detection rates. We first calculated the number of 2-week sampling periods (K) that would be needed to obtain a 90% chance of detecting a species at least once under current sampling design by using the detection probability (p) for each species in both summer and winter, by solving for K using:

$$0.90 \geq 1 - (1-p)^K$$

Second, we used a Student's t-test to determine if there was variation between detection rates for summer and winter. Last, we used linear regression to determine if the detection rates for a given carnivore species between the seasons were correlated.

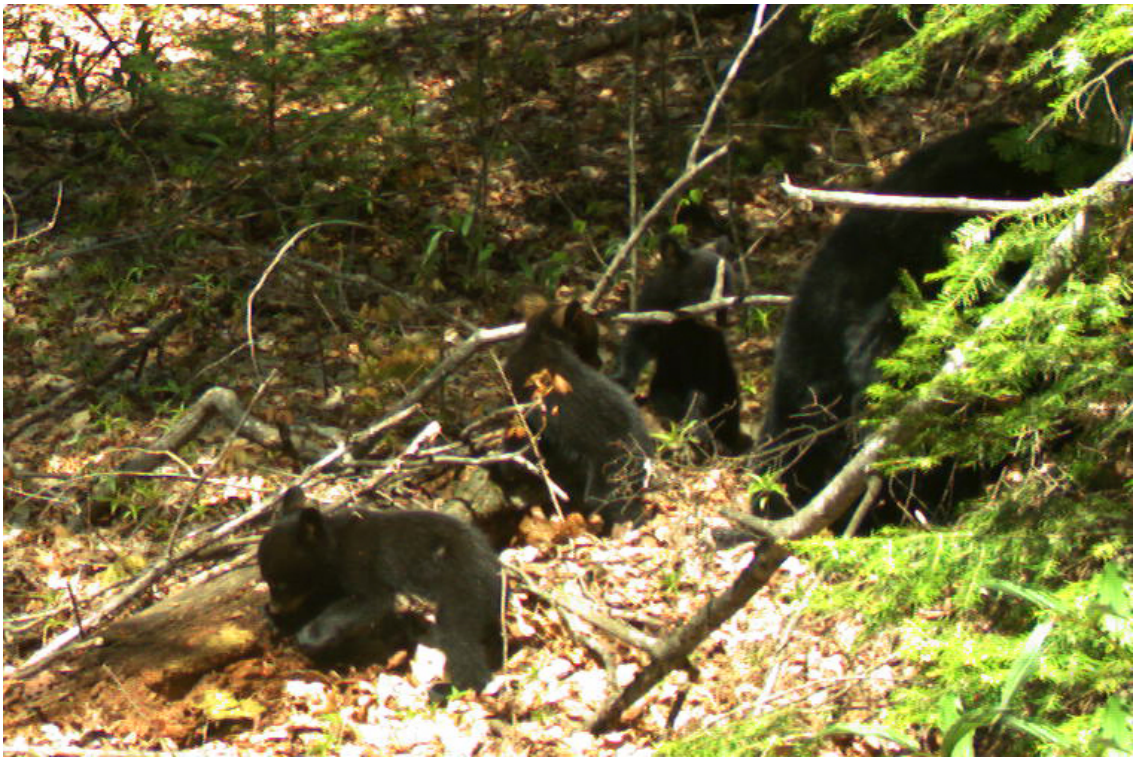

*Photo 4. A trio of black bear cubs with their mother on Bear Island.*

# RESULTS

*Summary Statistics*

We deployed 88 cameras on 13 islands, collecting 92,694 photographs (not including time-lapse photographs) across 18,721 trap nights. We documented 3,591 wildlife events, including 1,070 carnivore events. We had a mean of 6.6 (±2.2 SE) cameras per island (range 2-30), and our camera density averaged 1.23 (±0.19 SE, range 0.74-3.08) cameras/ km$^2$.

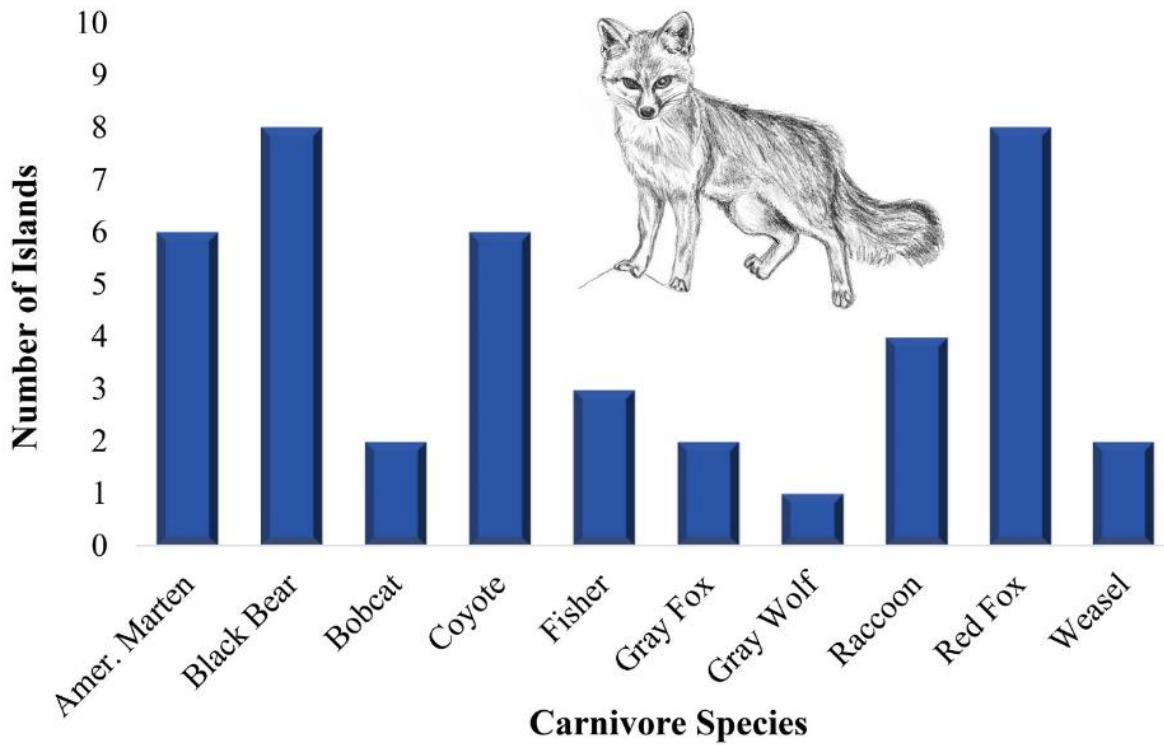

*Figure 3. The number of islands on which each terrestrial carnivore was documented (drawing by Yiwei Wang).*

*Camera Deployment Protocol*

We developed a standardized protocol for placing cameras to detect terrestrial carnivores (Appendix 1). We initially used our protocol to deploy cameras on Stockton Island. We then placed cameras on an additional 12 islands, and found the protocol was robust across islands of different sizes for detecting the presence of terrestrial carnivores. We detected 27 species including 10 terrestrial carnivores and one non-terrestrial carnivore (river otter, *Lontra canadensis*). Each of the terrestrial carnivores was found on 1 to 8 islands (Figure 3).

*Carnivore Species Richness*

The mean richness of carnivores on an island was 3.23, and the number of carnivores we detected varied by island from 0 (North Twin) to 10 (Stockton Island) (Figure 4).

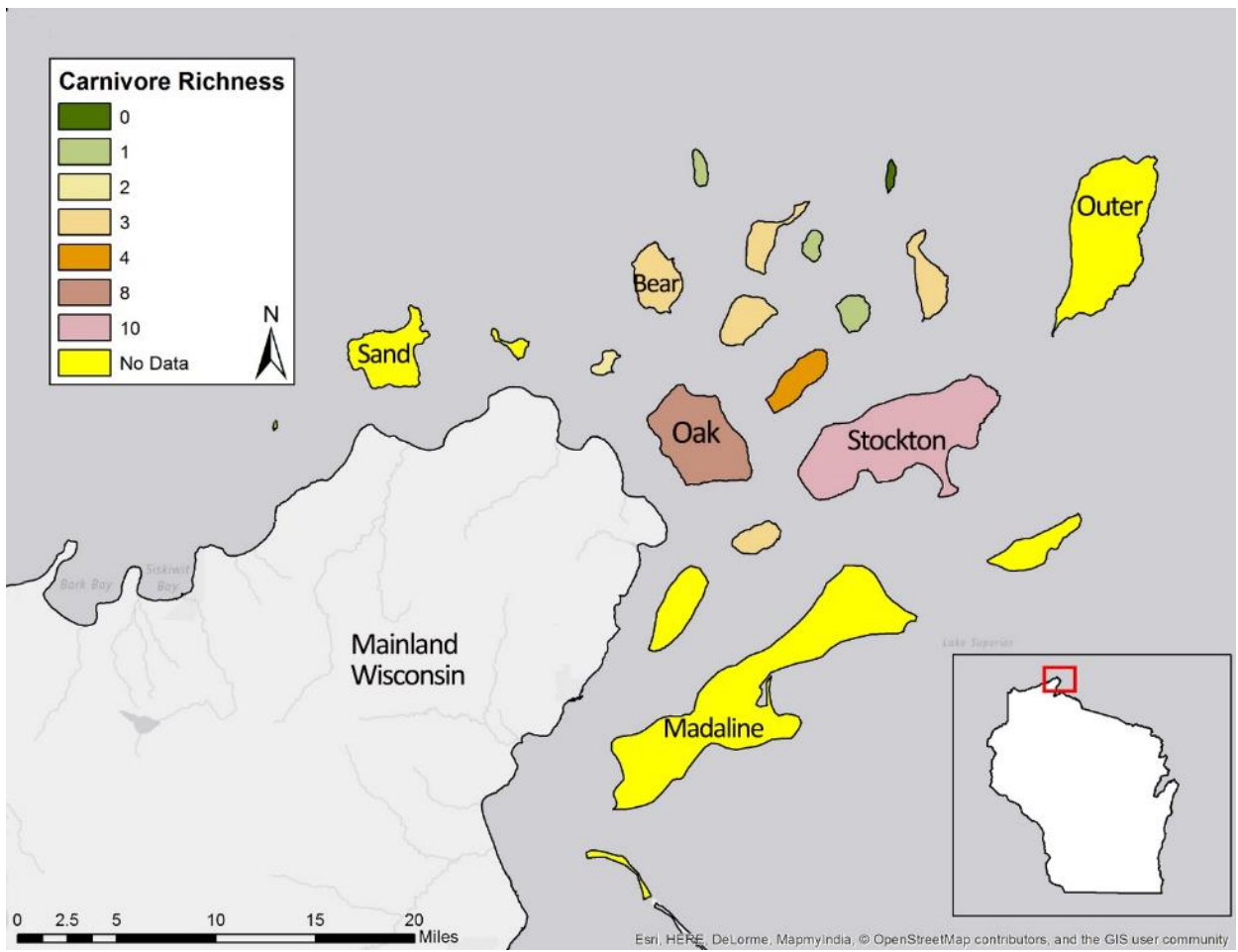

*Figure 4. Map of carnivore species richness for each of our study islands.*

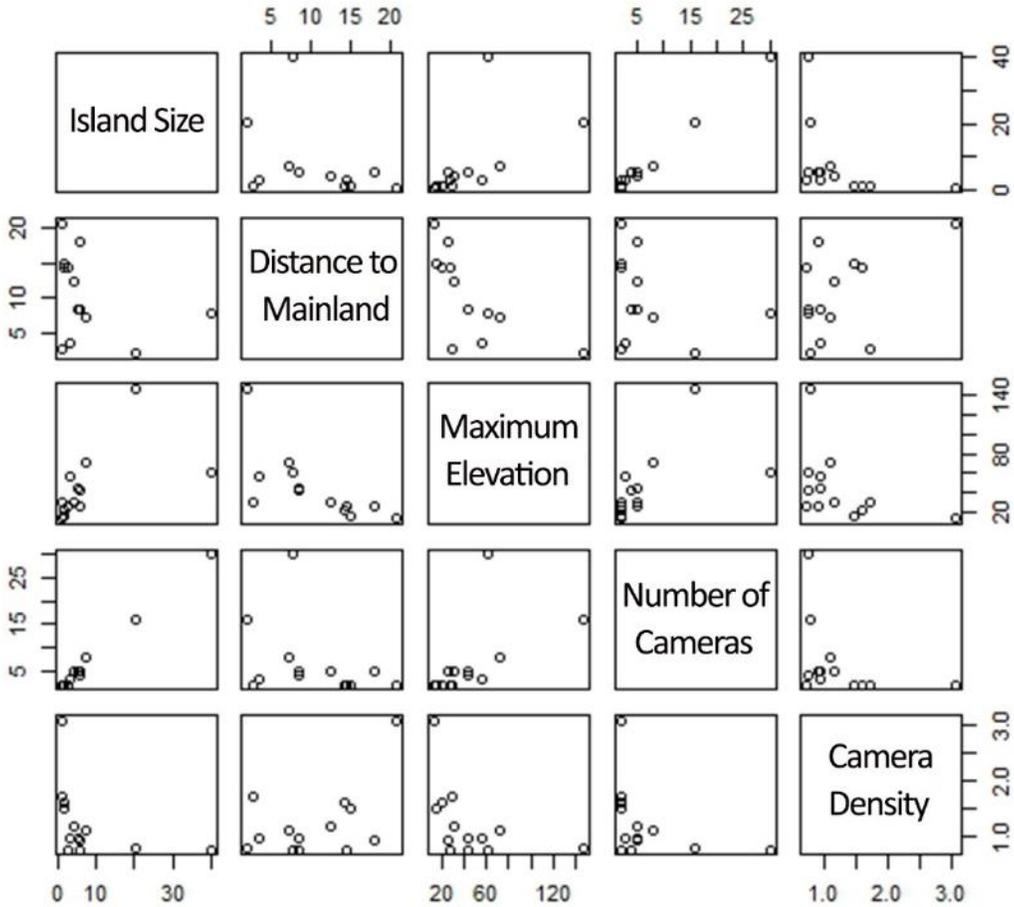

*Figure 5. Scatterplots of variables potentially driving carnivore species richness.*

The best single variable to explain carnivore richness on the Apostle Islands was island size ($F_{1,11} = 80.46$, $R^2 = 0.87$, $p < 0.0001$) (Figure 5). Distance to mainland ($F_{1,11} = 4.63$, $R^2 = 0.23$, $p = 0.0538$) and maximum elevation ($F_{1,11} = 12.13$, $R^2 = 0.48$, $p = 0.0051$) also had a significant relationship. The best model was island size and distance from mainland ($F_{2,10} = 73.91$, $R^2 = 0.92$, $p < 0.0001$) (Figure 6).

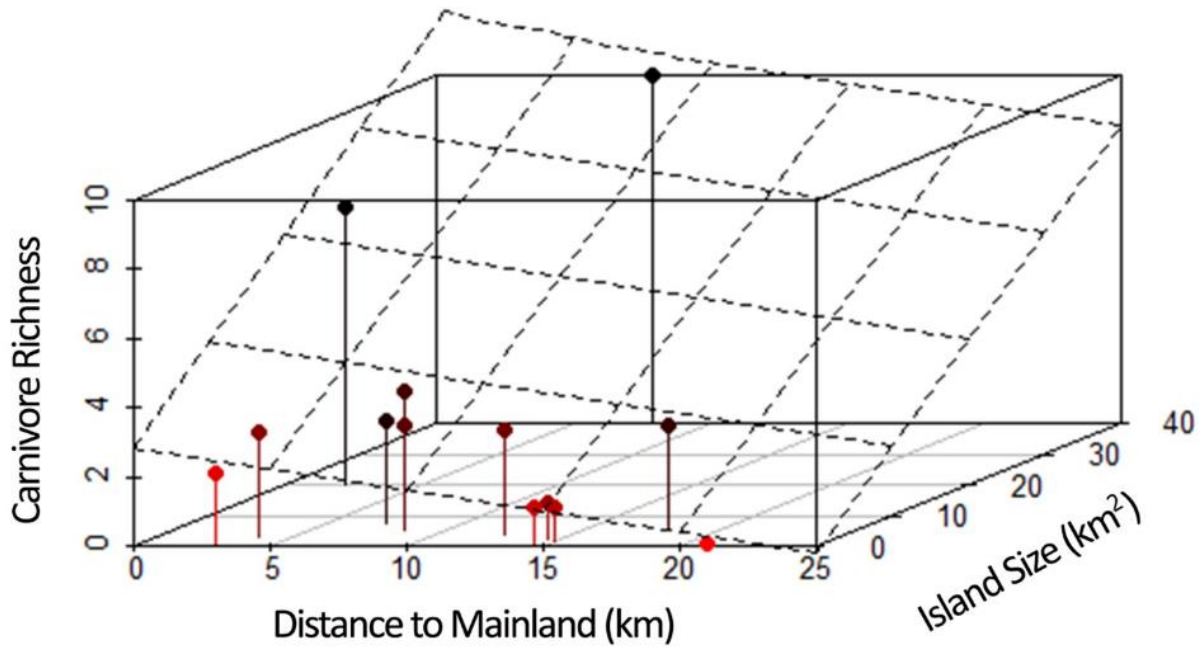

*Figure 6. The relationship of island size and distance to mainland with carnivore species richness on 13 islands in the Apostle Islands National Lakeshore.*

***Carnivore Relative Abundance and Distribution***

There was variation in the distribution and relative abundance of different carnivore species (Figures 7a-7j). There was a correlation for some species, including black bears which were distributed on a high number of islands (n = 8) (Figure 7b) and had a high relative abundance (2.64), and weasels which were distributed on a low number of islands (n = 2) (Figure 7j) and had a low relative abundance (0.1). Other species, such as red fox, were distributed on a high number of islands (n = 8) (Figure 7i) but had a low relative abundance (0.31).

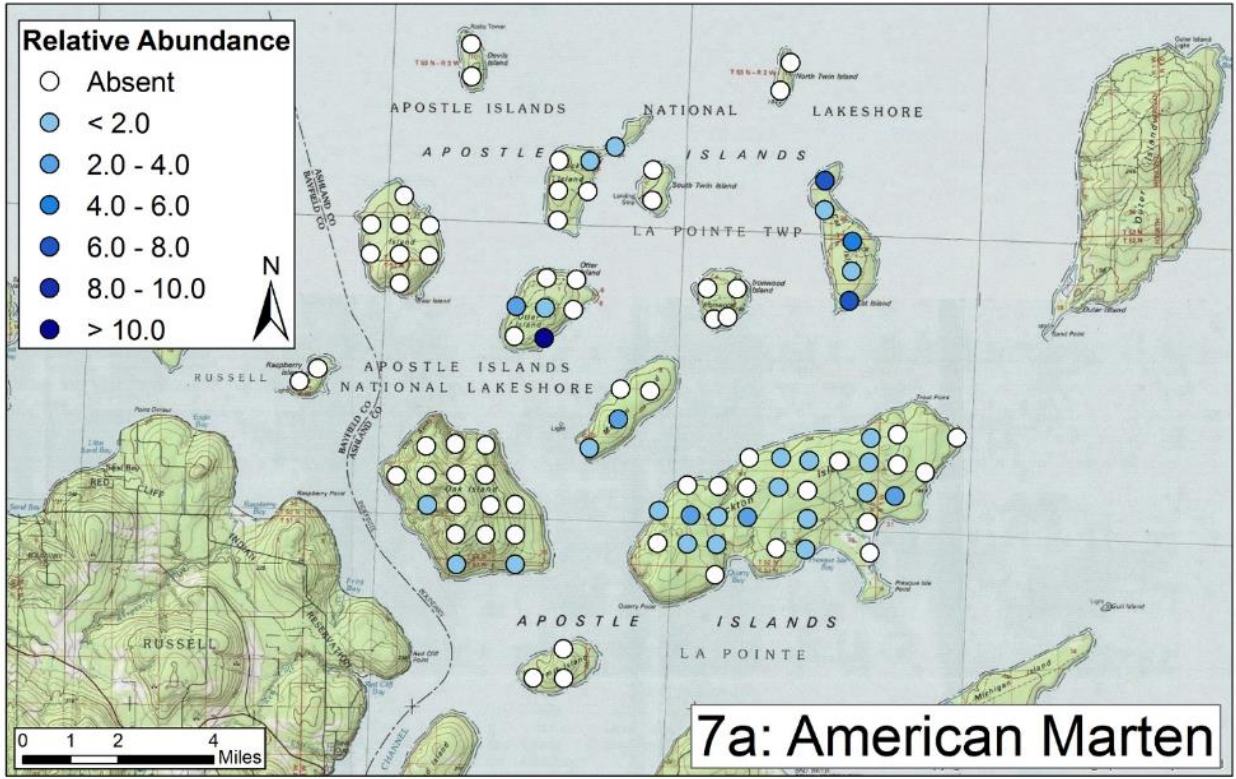

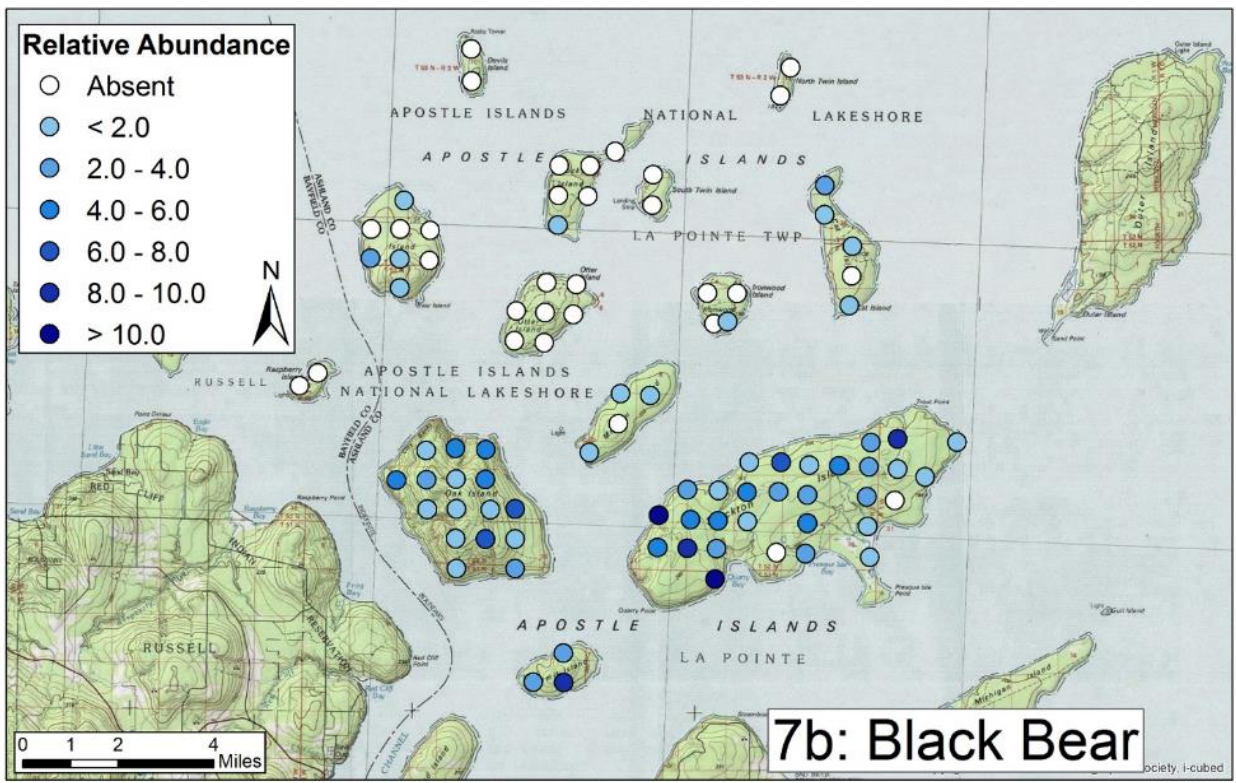

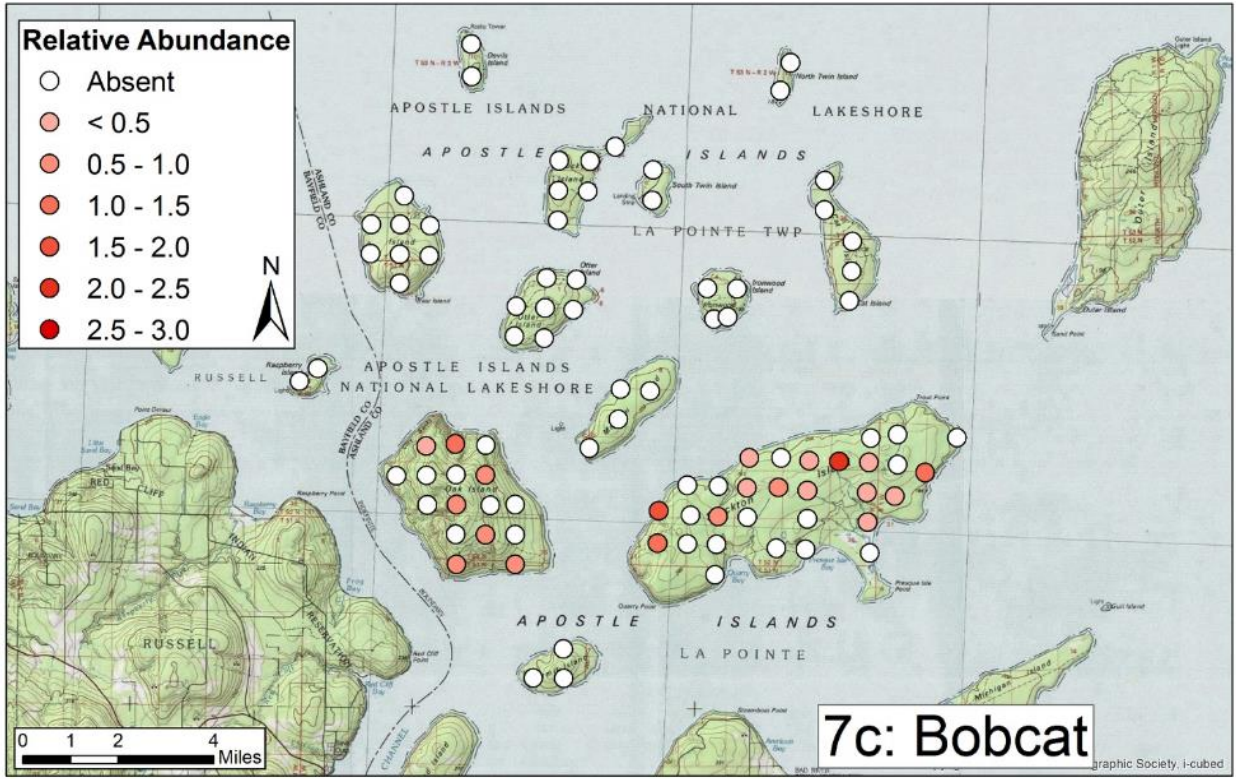

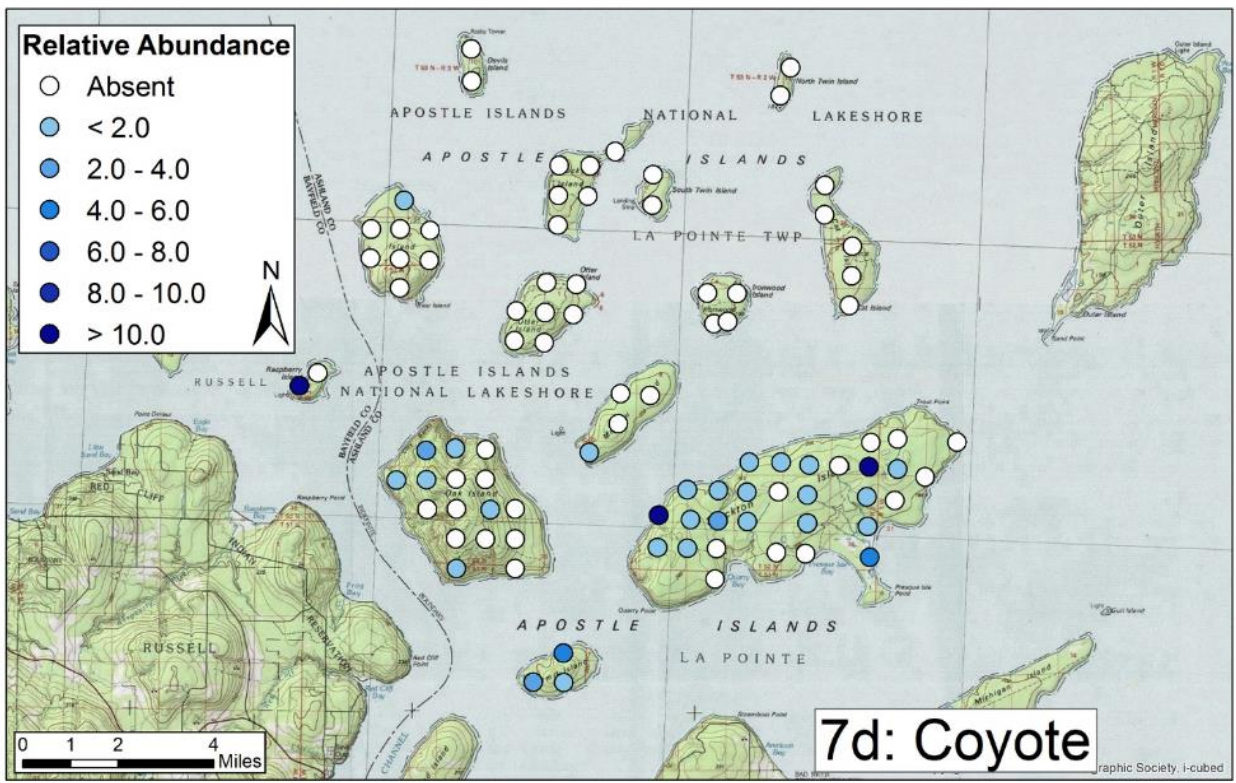

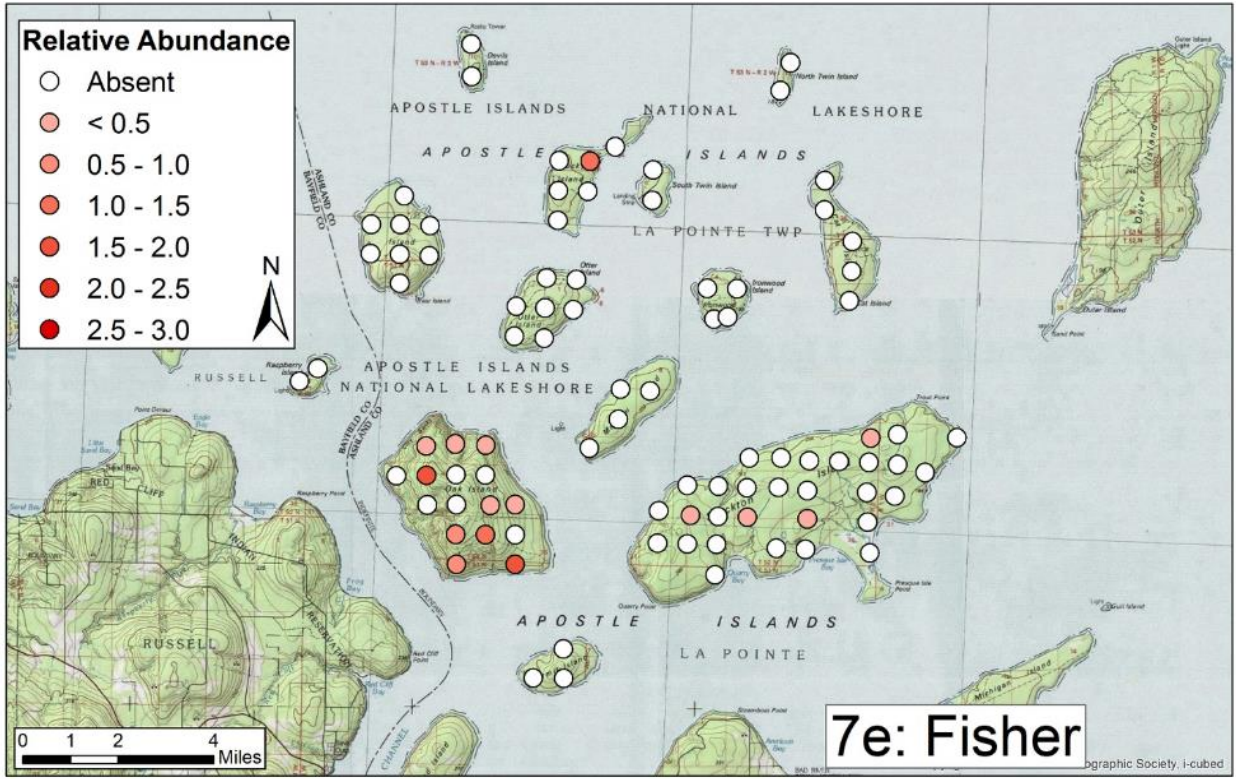

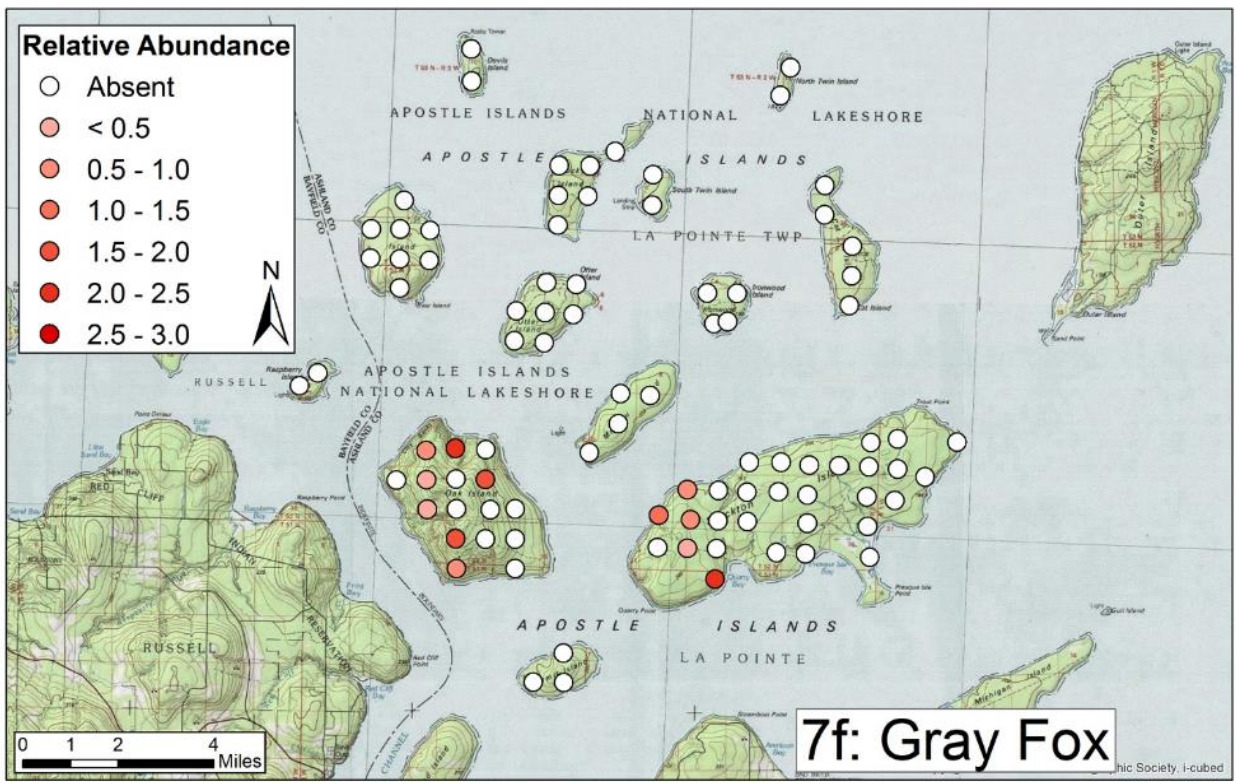

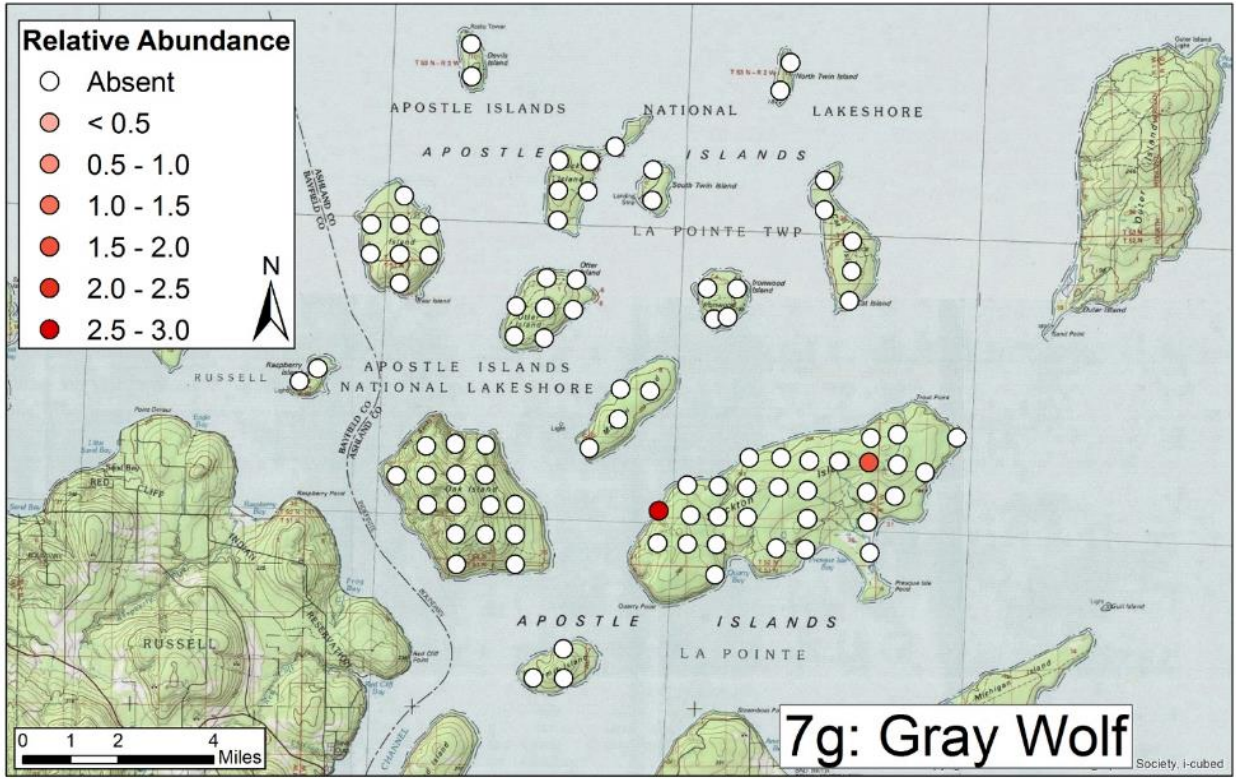

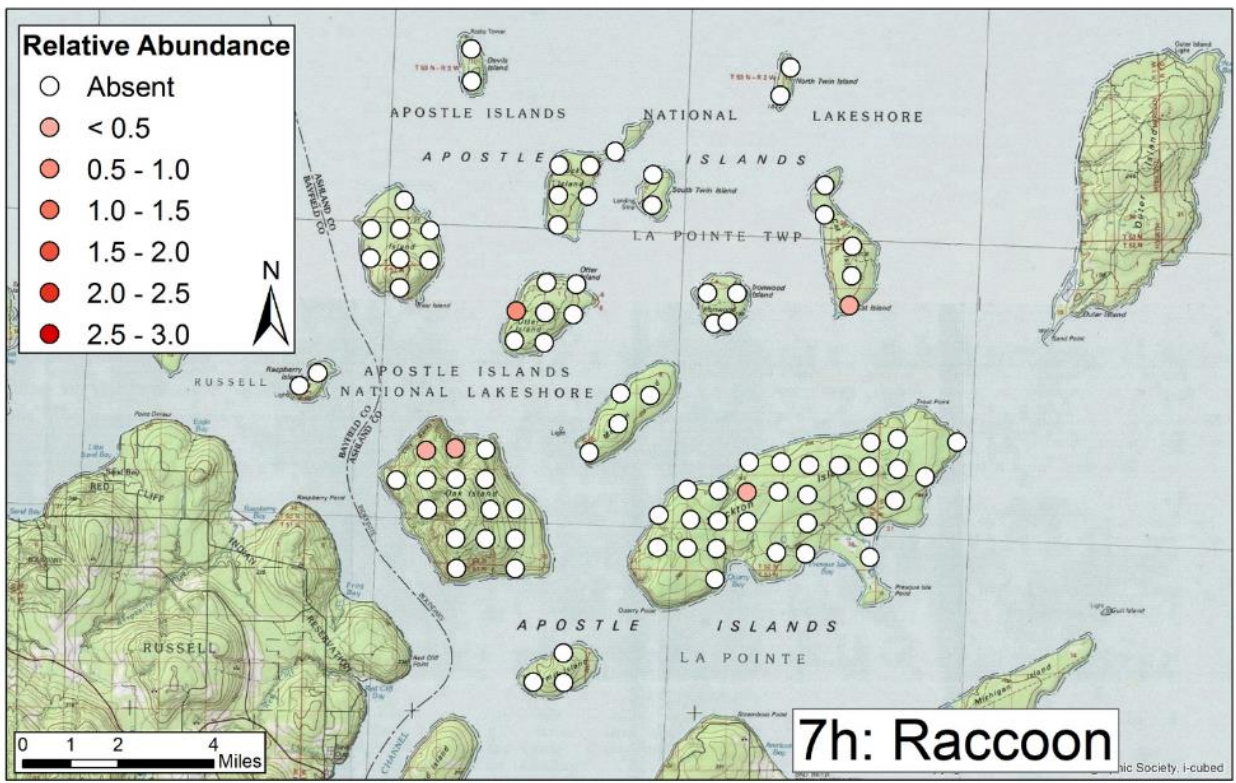

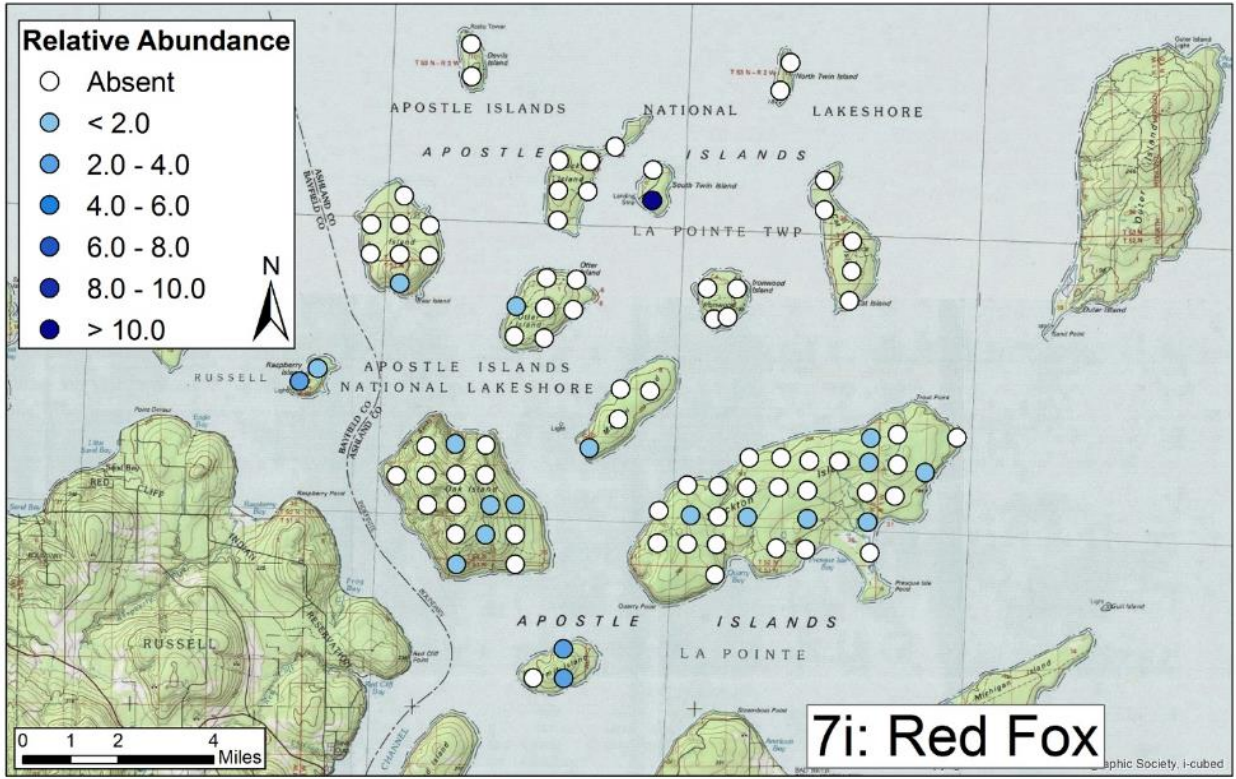

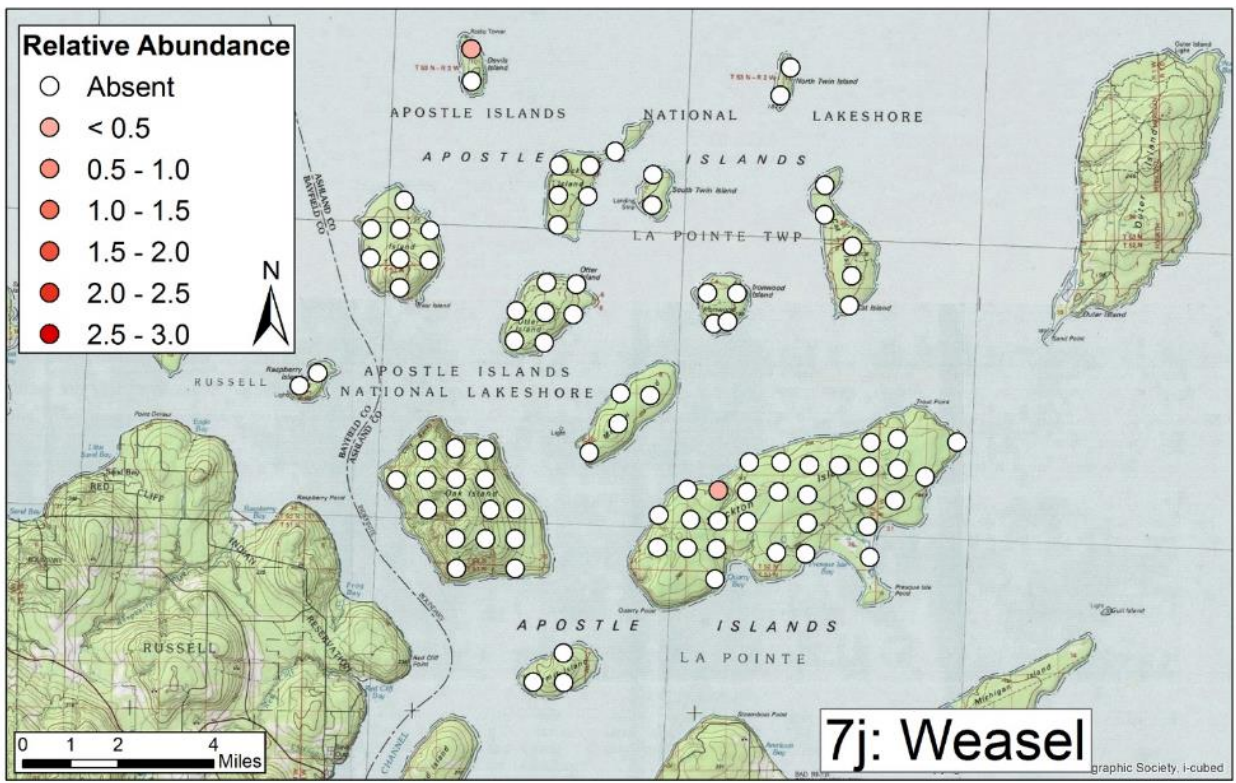

Relative abundances for carnivores ranged from a low of 0.01 (±0.01 SE) for weasels (*Mustela spp.*) to a high of 2.64 (±0.52 SE) for black bears (Figure 8). The relative abundance of a species had a significant marginal relationship with the number of islands on which they were found ($F_{1,8}$ = 8.55, $R^2$ = 0.52, $p$ = 0.0192).

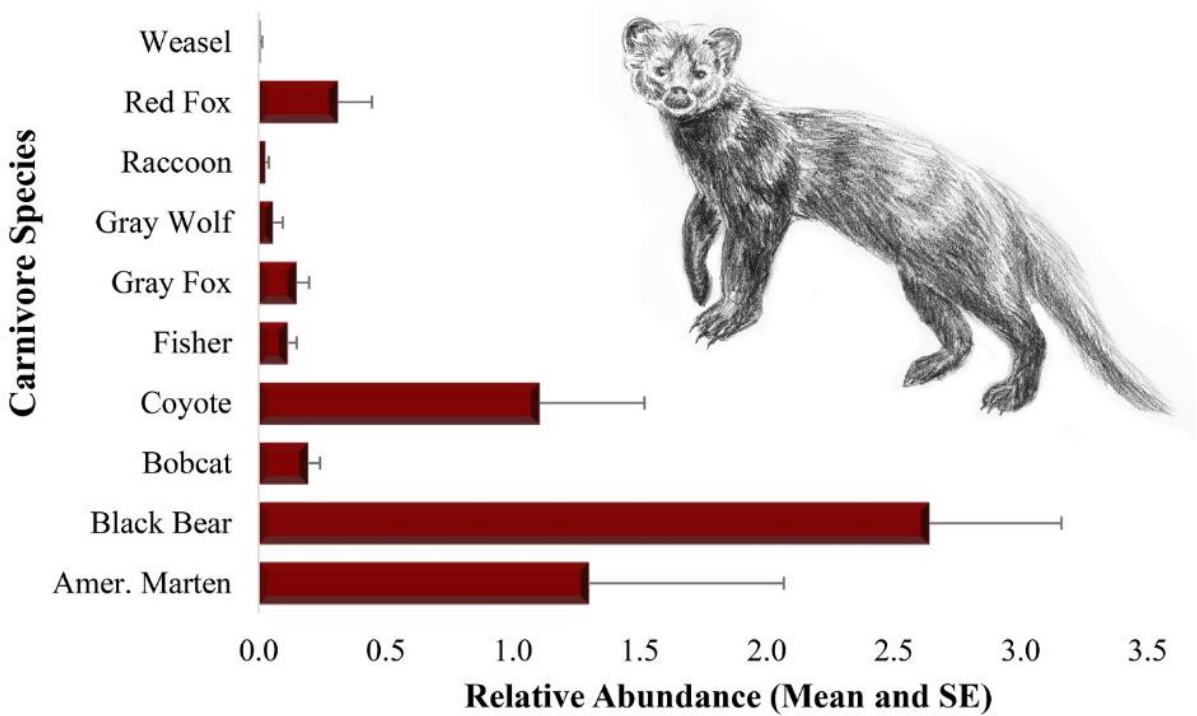

*Figure 8. Mean relative abundance and standard error for the 10 terrestrial carnivores detected in the Apostle Islands National Lakeshore (drawing by Yiwei Wang).*

### *Carnivore Occupancy and Detection Rates*

Carnivore occupancy ranged from lows of 0.09 for gray wolves (*Canis lupus*) and 0.11 for weasels to a high of 0.82 for black bears (Figure 9). Our detection rates for carnivores ranged from 0.38 for red fox to 0.88 for black bear in summer, and ranged from 0.01 for weasel to 0.20 for coyote in winter (Table 2). Detection rates were significantly higher in summer than winter ($t_{16}$ = 8.28, $p < 0.0001$), and the detection rates for species differed by season ($F_{1,8}$ = 6.82, $R^2$ = 0.49, $p$ = 0.0348).

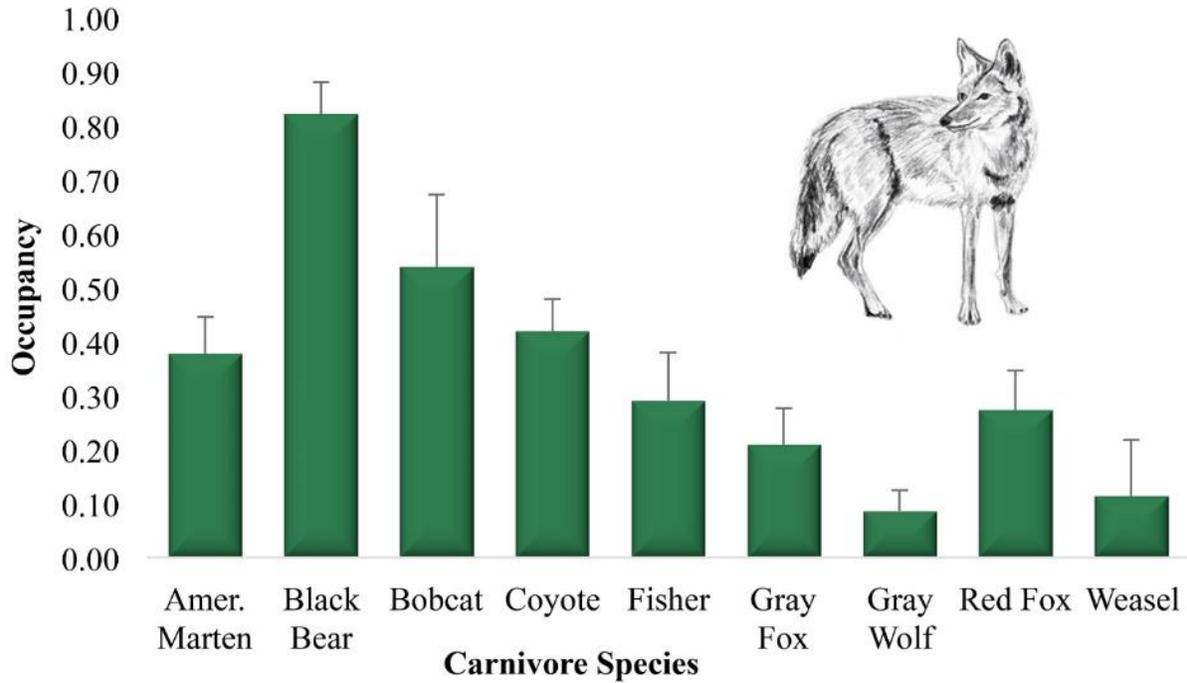

Figure 9. Occupancy for terrestrial carnivores we detected in APIS (drawing by Yiwei Wang).

Table 2. The detection rates of carnivores in the Apostle Islands National Lakeshore. We report the detection rates for summer and winter, our estimate of the weeks needed for monitoring to detect a species with 90% accuracy, and the detection rates for a full season (26 weeks) of monitoring (assumes the species is present and assumes camera densities used in this study).

|  | Detection Rate | | Weeks Needed | | Full Season | |
| --- | --- | --- | --- | --- | --- | --- |
| Species | Summer | Winter | Summer | Winter | Summer | Winter |
| American Marten | 0.40 | 0.12 | 10 | >26 | 1.00 | 0.80 |
| Black Bear | 0.88 | 0.08 | 4 | >26 | 1.00 | 0.65 |
| Bobcat | 0.75 | 0.03 | 4 | >26 | 1.00 | 0.28 |
| Coyote | 0.42 | 0.20 | 10 | 22 | 1.00 | 0.95 |
| Fisher | 0.71 | 0.04 | 4 | >26 | 1.00 | 0.38 |
| Gray Fox | 0.62 | 0.06 | 6 | >26 | 1.00 | 0.53 |
| Gray Wolf | 0.84 | 0.04 | 4 | >26 | 1.00 | 0.39 |
| Red Fox | 0.38 | 0.09 | 10 | >26 | 1.00 | 0.71 |
| Weasel | 0.79 | 0.01 | 4 | >26 | 1.00 | 0.14 |

Because of variation in detection rates between seasons, the number of weeks needed to detect each species with a 90% certainty was notably longer in winter than summer (Table 2). In fact, most species need greater than the entire winter season to achieve 90% certainty (Table 2).

Allen et al. 2016 – Apostle Island Carnivore Guild# DISCUSSION

## *The Carnivore Community*

We found a higher level of carnivore richness, abundance, and occupancy than we expected, and we consider our monitoring protocol to have worked exceptionally well for monitoring and describing the carnivore community in APIS. Before the study began we expected to document coyote and black bear, but considered other species including gray wolf, red fox, fisher (*Pekania pennanti*) potentially present. Although we could not identify weasel detections to species, the 10 terrestrial carnivores we documented potentially represent all but two of the native terrestrial carnivores present in Wisconsin, (exceptions: American Badgers (*Taxidea taxus*) and striped skunks (*Mephitis mephitis*)). There was a noticeable trend of absence (striped skunk) or low abundance (raccoon) of synanthropic carnivore species, as well as the absence of Virgina opossums (*Didelphius virginianus*), a synanthropic mesopredator. Low levels of human development and recreation in APIS may play a role in supporting carnivore species (such as fishers, bobcats, *Lynx rufus*; and gray wolves) that would otherwise avoid human disturbance (Haskell et. al 2013).

One of the most important outcomes of our study was understanding how island biogeography theory (e.g., MacArthur and Wilson 1967, Wilson 2009) may explain carnivore species richness. Size of an island had a strong positive correlation with carnivore richness, while distance to mainland Wisconsin had a negative relationship; and the combination of the two variables had a very strong correlation with carnivore richness ($R^2 = 0.92$, Figure 5). Size of an island likely dictates the diversity and abundance of resources, such as prey and habitat, which are available for carnivores. Prey abundance has been found to reduce avoidance behavior between sympatric and otherwise antagonistic carnivores (Grassel et al. 2015). A larger diversity of habitat

Page | 27

also allows carnivores that compete for the same resources to establish foraging or behavioral niches to partition resources (Lesmeister et al. 2015, Wang et al. 2015). It is therefore not surprising that larger islands support more carnivore species. The negative effect of distance to mainland Wisconsin suggests that the populations of a given carnivore species may be dependent on periodic influxes from the mainland to maintain their population. This may be because in an island system, the end of the archipelago acts as a geographical limit to the dispersal of young animals, or islands far from the mainland may act as population sinks. Long-term monitoring would help elucidate trends in carnivore community dynamics on the islands and reveal whether the current diversity of carnivores is sustainable. It appears that APIS is a model system for studying the effects of island biogeography on the carnivore community, and we encourage future ecological studies both in APIS and in archipelagos in the Great Lakes Region.

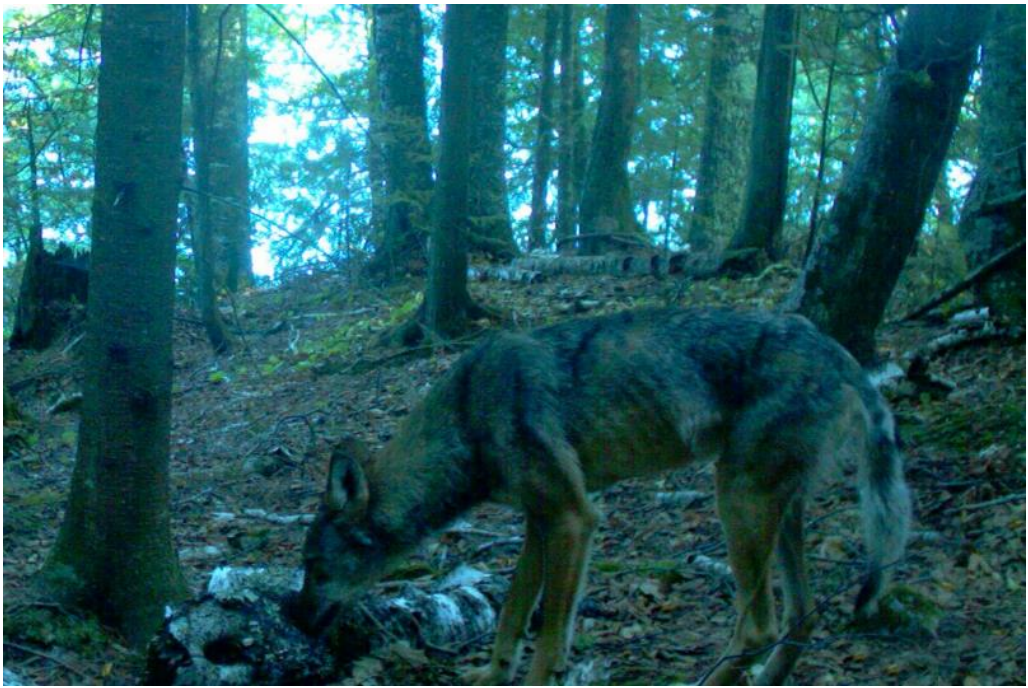

*Photo 5. A surprising detection of a gray wolf on Stockton Island.*

None of the islands in the archipelago are likely large enough to sustain populations of mammalian carnivores in the face of demographic stochasticity or the genetic effects of small population size. Hence, one important area for future study is determining how carnivores colonize and move between islands. In the Apostle Islands, the movement of mammals between the islands and outside of the archipelago is likely through either swimming or travel across ice in winter, and community dynamics may therefore be mediated by the effects of a warming climate. Reports exist of some mammal species swimming long distances between islands and the mainland (Jackson 1920, Wilton et al. 2015), but the role of long-distance swimming as a movement strategy for terrestrial carnivores is poorly understood. Alternatively, species can immigrate and emigrate from the archipelago in winter, when ice forms connective bridges between the islands and the mainland. For example, gray wolves originally populated Isle Royale National Park in the late 1940s by crossing an ice bridge connecting the island to mainland Ontario (Adams et al. 2011). Limnologists have documented declines in the duration of lake ice in the northern hemisphere over the last 150 years (Magnuson et al. 2000), suggesting that if travel across the ice is the primary mode of recolonization it may be affected by climate change. This may in turn change the dynamics of the carnivore community in APIS, as some species may be dependent on ice for travel (e.g., wolves, coyotes and red foxes) while others may not be (e.g., black bears).

Abundance of prey may be an important aspect of sustaining the carnivore community. The carnivore species documented appear to outnumber the identified potential prey species and this raises questions about what sustains the larger carnivores. Red squirrels (*Tamiasciurus hudsonicus*) seem abundant, while other rodents and hares were less frequently documented on camera. However, cameras are more likely to be triggered by larger animals, and cameras deployed as in this study may not be an effective method of documenting prey populations. Implementing small-mammal trapping or other measures may inform our understanding of carnivore community dynamics and competition. Deer present a more substantial food source for gray wolves and coyotes (Arjo et al. 2002, DelGiudice et al. 2009), as well as the rest of the carnivore community when acting as scavengers (DeVault et al. 2003, Allen et al. 2015). Deer populations may be an important aspect of carnivore diet on the islands, but further study of carnivore diets is needed to understand if prey availability affects carnivore populations over time in APIS.

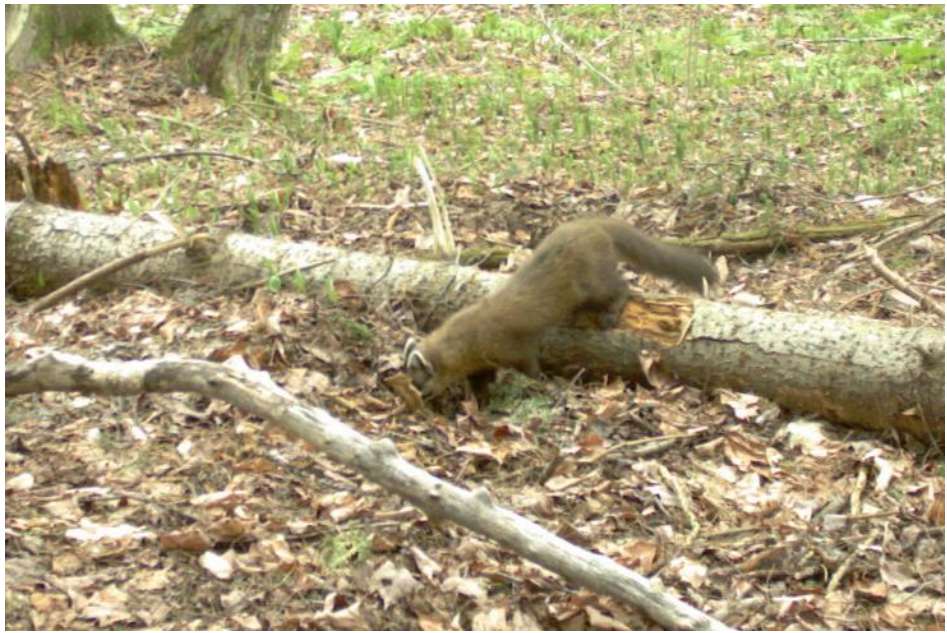

*Photo 6. An example of our surprising documentation of American martens in APIS*

One observation of particular interest was our detection of American martens in APIS. Martens were extirpated from the state of Wisconsin in the 1920s (Williams et al. 2007), and after reintroduction efforts on Stockton Island in the 1950s (Williams et al. 2007), martens had not been documented formally in APIS until this study. Because no focused survey efforts have been conducted on Stockton Island until this camera trapping project, it is difficult to confirm the source of the current martens inhabiting the island. It is possible that the martens are descendants of the 1950s reintroductions, or that they have naturally colonized from reintroductions on the mainland.

Allen et al. 2016 – Apostle Island Carnivore GuildThere is also the remote possibility that APIS martens are a relict population that survived the extirpation of martens on mainland Wisconsin. Martens had the second highest relative abundance among our carnivore species, which is surprising considering that martens are the only mammal listed as state-endangered in Wisconsin. The marten population in APIS may have important implications for the conservation of the mainland population, and it is therefore important to try and understand the factors that drive their distribution and abundance in APIS.

When looking at our results, it is important to consider the full spectrum of distribution, abundance, and occupancy to understand a carnivore species in APIS. Some species, such as black bears, had a high abundance and occupancy, and were distributed on all but the smallest islands. In contrast, red foxes were found on many islands, but usually at low relative densities. We were surprised by our detections of American martens and gray wolves, but martens were widely distributed, while gray wolves were only found on the largest island. It is only when considering each of our measurements that we truly understand the patterns of the carnivore species on the islands. An important area of future research will be to determine how carnivore species are affected by the rest of the carnivore species in the community (e.g., Lesmeister et al. 2015, Wang et al. 2015).

*Camera Monitoring Protocol*

Our camera monitoring protocol was very successful for monitoring carnivore richness and distribution in APIS. Using these methods we documented a suite of terrestrial carnivore species, along with 17 other non-target species. The wealth of data collected has allowed us to use the data from our camera grid to calculate a variety of measurements, including species richness, relative abundance, distribution, detection probabilities, and occupancy. Using these methods long-term will allow APIS to document trends in carnivore richness and distribution. Data from camera grids

Page | 31

can also be used for other analyses, from population estimates to animal behavior and ecological dynamics. These data are particularly well-suited to population estimates using n-mixture models (Royle 2004) or spatially-explicit capture recapture (SECR) models (Chandler and Royle 2013), and creating population estimates of carnivores in APIS could be a valuable follow-up study.

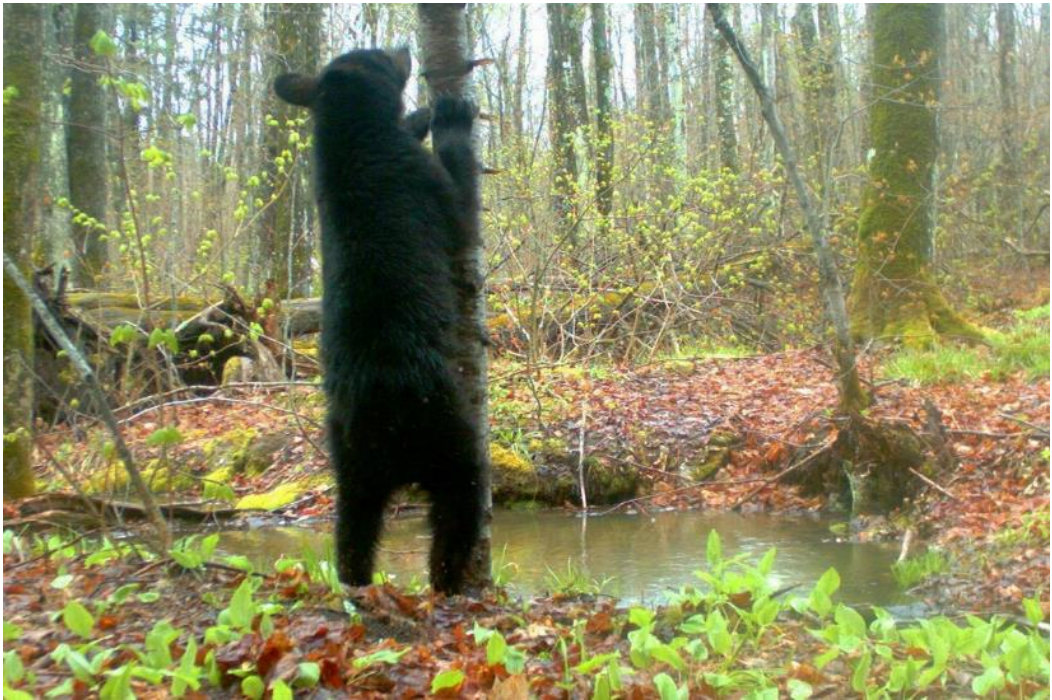

*Photo 7. Camera traps also document behaviors such as this black bear scent marking a tree.*

When creating a sampling grid for other ecosystems and species, it is important to consider our methodologies, but not necessarily the specifics we used. In our experience a 1x1 km grid worked well for the terrestrial carnivore guild of APIS; however, this should be tailored to a given set of target species, and study goals. For example, our protocol was designed for terrestrial carnivores and other large mammals, but if one were attempting to document semi-aquatic mammals one would likely use different sampling methods. In that case a modified linear grid that follows the shoreline of rivers, lakes, and wetlands would be more effective, as well as aiming the cameras towards water rather than toward openings in the forest, and potentially diminished detection probability should be accounted for (see Lerone et al. 2015, Evans 2017). Also, if targeting a specific terrestrial carnivore rather than the community as a whole, the grid density could be increased or focused on particular habitat or features. Similarly, if targeting smaller or less active species, the cameras may need to be set out for longer times or at higher density.

A key goal for developing our protocol was to obtain estimates of detection probabilities for each carnivore species to understand how long cameras need to be deployed to ensure detection (assuming that the species is there) and to determine the season of greatest detection probability. Our calculated detection rates may not be accurate for some species (gray wolf and weasels) because of their low number of absolute detections, and should be interpreted with caution at this time. For all species, our detection rates were notably lower in summer than in winter, suggesting that if detecting the presence of a carnivore is the only goal, that summer should be the season of deployment. Our winter monitoring periods were not evenly distributed across islands, however, and additional data will allow us to determine with greater accuracy how season affects detection rates. For APIS, we consider winter monitoring important to allow future analyses to determine colonization dynamics relative to ice linkages between islands being present, and if this changes the dynamics of the carnivore community.

Our low detection rates in winter may also be caused by low temperatures which diminish battery performance and reduce the area effectively monitored. A key component of the methodology we developed was to program cameras to take time-lapse photos each day (Appendix 1). These images allowed us to determine if a camera was functioning or not on a daily scale. One change that we would make is to program the time-lapse photos to be taken during the coldest part of the night (for example 2:00-3:00 am) to ensure camera functionality at the times that they are most likely to be affected by temperature.

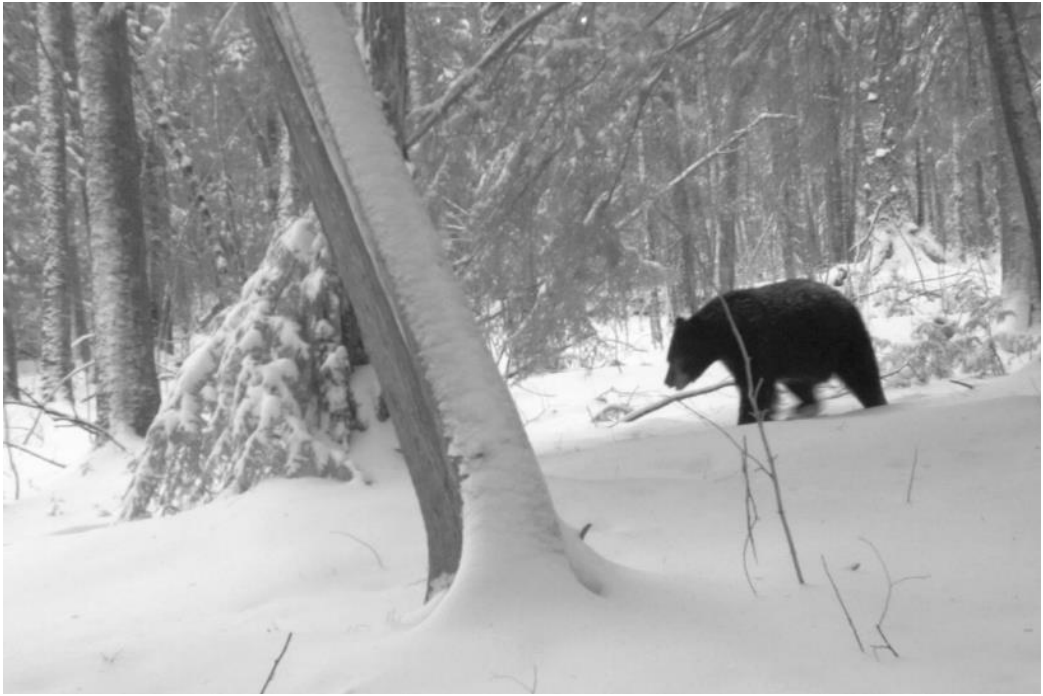

*Photo 8. An unusual winter detection of a black bear*

## Conclusions

The remarkable diversity and abundance of carnivores detected in the APIS archipelago exceeded our expectations and created an opportunity to investigate multiple aspects of their ecology. The camera traps collected information on species ranging in size from weasels to black bears, and relative abundances from single-camera detections to near-ubiquitous. Our methodology appears to be an effective approach to monitoring across such diverse criteria, and the wealth of data produced can inform park management as well as broader wildlife issues. The surprising detection of a population of American martens in the park may be of particular interest. As the only endangered mammal in Wisconsin, efforts to reintroduce and augment martens have been underway for over 60 years (Williams et al. 2007), but have often had limited success (Williams et al. 2007, Carlson et al. 2014), and understanding martens in APIS could provide valuable information for other conservation projects in Wisconsin.

Advances in basic and applied science often follow technological innovations. Remotely triggered cameras that are rugged enough for long-term monitoring under extreme environmental conditions are a relatively recent advance and the statistical and data-management techniques for optimizing large databases of camera-generated information are also currently evolving. These advances will enable more precise resolution of ecological patterns at finer temporal and spatial scales with reduced costs for labor and material. Managers should plan to exploit these trends to deal with emergent threats to the ecological integrity of protected areas such as those posed by climate change, increasing human impacts, invasive species, and reduced connectivity.

This project described the successful use of remotely-triggered cameras to monitor large-mammals (carnivores). In some sense, this effort represents one of the "low-hanging fruit" in terms of a fuller understanding of the landscape and terrestrial community dynamics of the APIS. Greater understanding of APIS ecology will require on-going monitoring of carnivores to evaluate temporal dynamics as well as related ecological evaluations (e.g. small mammal dynamics, plant community dynamics) to understand trophic effects.

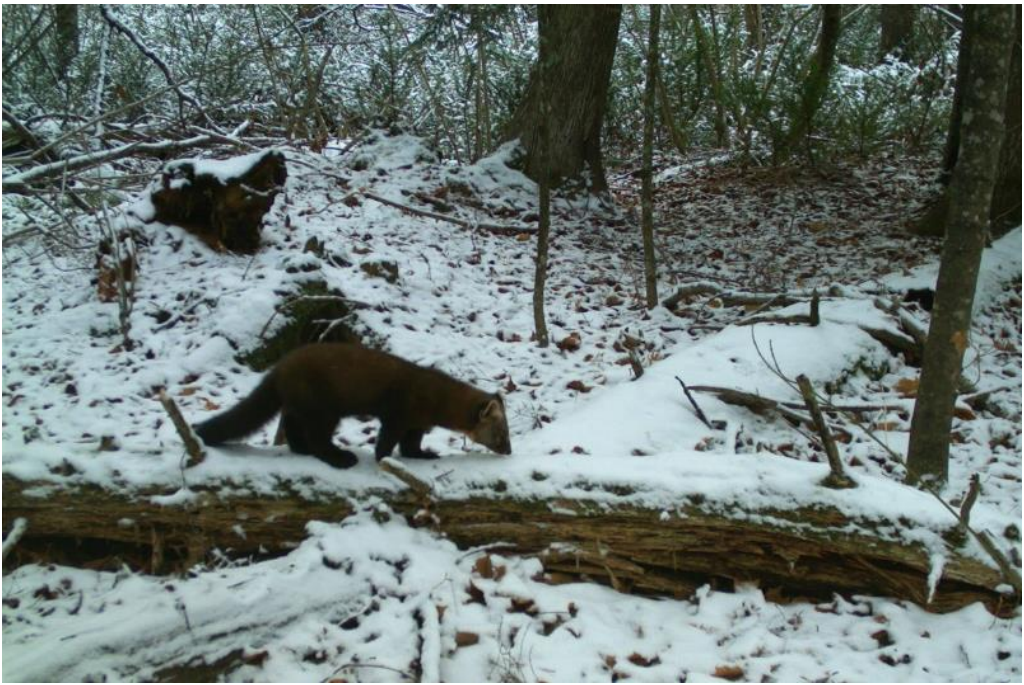

*Photo 9. An American marten in a winter forest.*

Allen et al. 2016 – Apostle Island Carnivore Guild

Allen et al. 2016 – Apostle Island Carnivore Guild

# Appendix 1
# Protocol for Creating and Setting a Camera Grid

# From:
# Survey techniques, detection probabilities, and the relative abundance of the carnivore guild on the Apostle Islands (2014-2016)

# Final Report
**December 15, 2016**

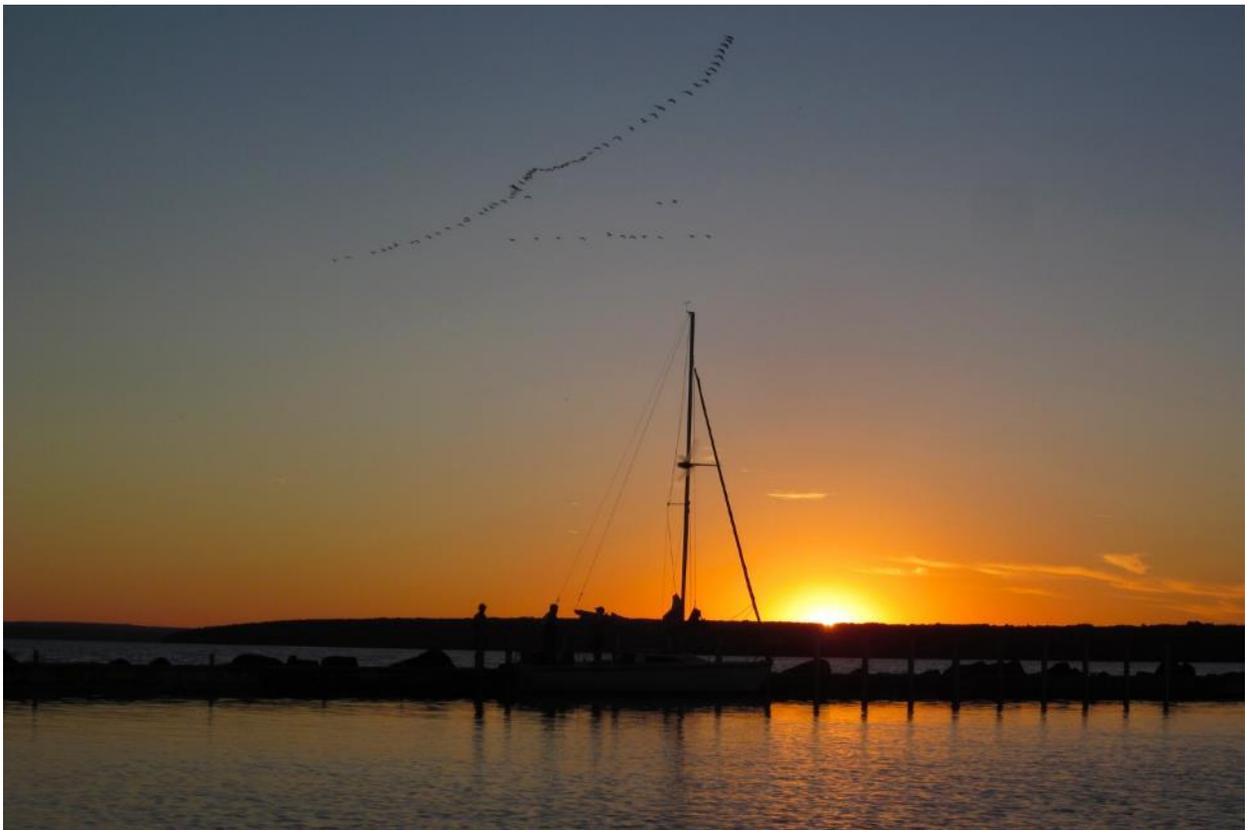

Photo by Bryn Evans

Suggested Citation:
Allen, M. L., B. E. Evans, M.E. Wheeler, M. A. Mueller, K. Pemble, E. R. Olson, J. Van Stappen, and T. R. Van Deelen. 2016. Survey techniques, detection probabilities, and the relative abundance of the carnivore guild on the Apostle Islands (2014-2016). Final Report to the National Park Service.

*Creating a Sampling Grid for Camera Deployment*

Effective camera trap designs for monitoring wildlife depend on sampling in a standardized and rigorous method. One of the most effective designs for monitoring is to create a sampling grid on which to place cameras because it enables both systematic (sample all grid cells) and systematic-random (sample a randomly chosen subset of grid cells) designs for unbiased camera placement. The density of cameras in the grid used is dependent on both the monitoring goals and logistical constraints of a project. The spacing of the cameras is dependent on the detection probabilities and home range size of the target species.

To meet the goals of this project we overlaid satellite imagery of the APIS area with a 1km$^2$ grid. We then created a 'camera deployment location' at the center of each grid that had >50% of its surface area on land (Figure 9). This allowed standardized (even and repeatable) camera placement across all islands, at a scale that we thought would be rigorous enough to meet the detection and abundance measurement goals of our project based on previous experience with an earlier study on Sand Island (Bartnick, et al. 2013).

Different models and brands of cameras vary in their performance. To ensure standardized monitoring we used the same model of cameras for all of our monitoring. For the study we used RECONYX HC600 Hyperfire™ High Output Covert infrared digital game camera (RECONYX, Inc., Holmen, Wisc., USA).

*Camera Placement*

We created standardized and documented procedures for placing cameras in strategic locations. This allowed for multiple field staff, with varying levels of skill, to set cameras in a similar pattern and maximize the detection of our target species (e.g., O'Connell et al 2011).

We created a systematic naming system for each of our camera deployment locations for consistency and accuracy in tracking the data. Our naming system included: A two-letter notation for the project ("CN" for the Carnivore project), the two first letters of the island, a two-digit number unique to the camera deployment location and whether it had lure or not (presence or absence of L). For example:

CNOA01L = Carnivore Project, Oak Island, Camera Deployment Location 1, lure used

CNST14 = Carnivore Project, Stockton Island, Camera Deployment Location 14, no lure used

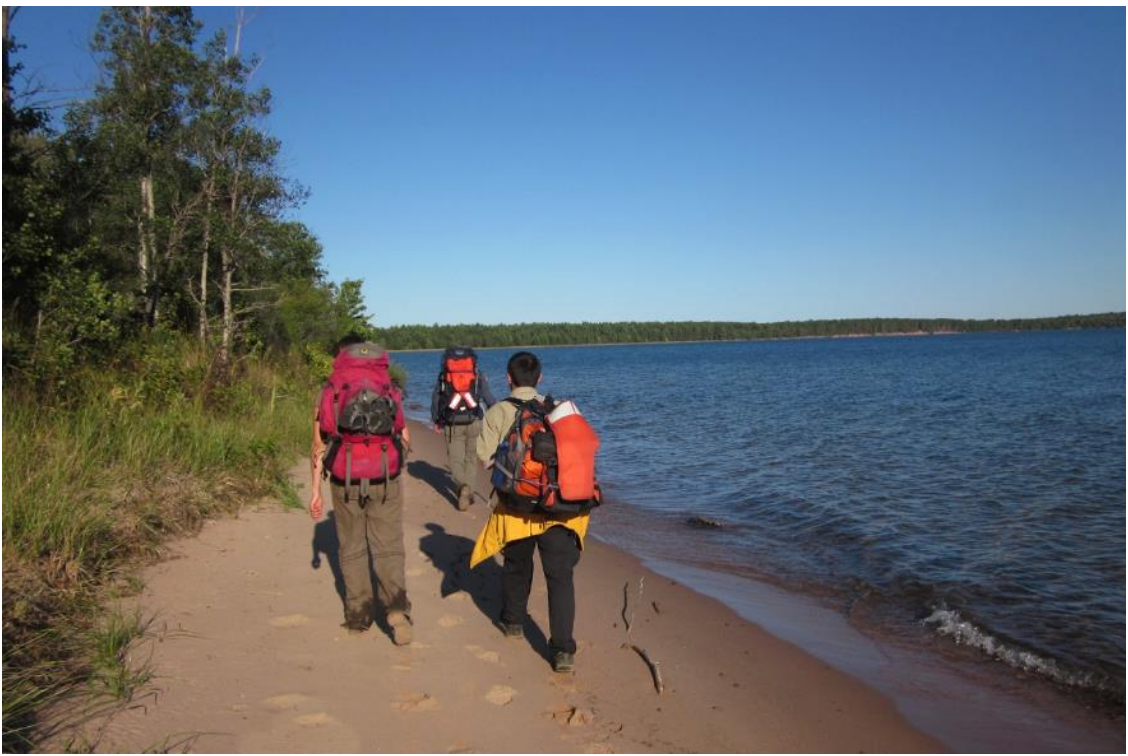

*Photo 10. Our field staff hiking in to place cameras on Stockton Island.*

We split our camera placement procedure into 6 steps:

1) *Selection of camera location*

We navigated to our target camera deployment location (the center of a grid cell) using handheld GPS units. Upon arriving we walked in concentric circles until we intersected animal sign, a clearing, a trail, or other site relatively free of obstructions that would serve as a natural travel route for a large mammal. Most suitable high-use areas were characterized by a small opening in the vegetative cover where ≥2 trails intersected, and included trees for camera placement, and adequate space (≥3 m, ≤7 m) between a suitable tree and the trail intersection. We suspect several target carnivores are attracted to large, downed woody debris, and therefore we prioritized sites with logs or stumps if available. Generally, this allowed for ideal placement of cameras and reduced the amount of vegetation in front of cameras that could cause false triggers or obscure the camera images.

2) *Programing of camera settings*

It is important camera settings are the same to ensure even and accurate sampling across all sites. In order to do this, we turned the camera on, entered the security code (if applicable) and then set the appropriate settings. The following is an example of how we set the cameras for the RECONYX HC600 Hyperfire™ High Output Covert infrared digital game camera (RECONYX, Inc., Holmen, Wisc., USA):

a) Change Setup → OK; Advanced → OK; Trigger → OK; Motion Sensor → ON; Sensitivity → HIGH; Pics per Trigger → 3; Picture Interval → RAPIDFIRE; Quiet Period → NO; Finished → OK (This sequence sets the motion sensor on with the high sensitivity and programs the camera to take 3 photos in rapid sequence after each trigger)

b) Change Setup → OK; Advanced → OK; Time Lapse → OK; AM Period → ON; AM Start → 11:00 AM; AM End → 12:00 PM; PM Period → OFF; Picture Interval → 1 HOUR; Finished → OK (This sequence programs the camera to take a single photo each day at 11 am)

 c) Change Setup → OK; Advanced → OK; Night Mode → OK; High Quality → ON; Finished → OK (This sequence programs the camera to operate at night using infra-red illumination)

 d) Change Setup → OK; Advanced → OK; Date/Time/Temp > OK; Finished → OK (This sequence programs the camera to record date, time and temperature with each image)

 e) ARM CAMERA → OK (This sequence programs the camera to begin monitoring)

**The red light on camera front should flash, and the camera is now ready to be deployed.**

Optional steps include:

 f) Change Setup → OK; Advanced → OK; User Label > OK; Choose Add OR View/Change, and enter your unique user label; Finished → OK (This sequence adds a unique user label for all of the images taken)

 g) Change Setup → OK; Battery Type →Lithium; Finished → OK (This sequence optimizes camera function for different battery types)

3) *Deploy the camera*

We affixed cameras to a tree approximately 0.75-1.5 m above the ground level and removed any foreground obstructions (e.g., long grass, shrubs, saplings, and branches) within the field of view of the camera. We placed cameras 3-5 m from the trail intersection, and faced the cameras at a 45-degree angle in relation to the most traveled section of trail. A small stick could be placed behind the camera to provide a downward angle and ensure the camera captures the entire area you are trying to survey. It was important make sure there were no large trees or objects in the main field of view of the camera, as this could adversely affect motion detection and nighttime flash range and light balance of the photo. We avoided setting cameras in a location where branches or vegetation of any kind could block the camera or grow in front of the camera.

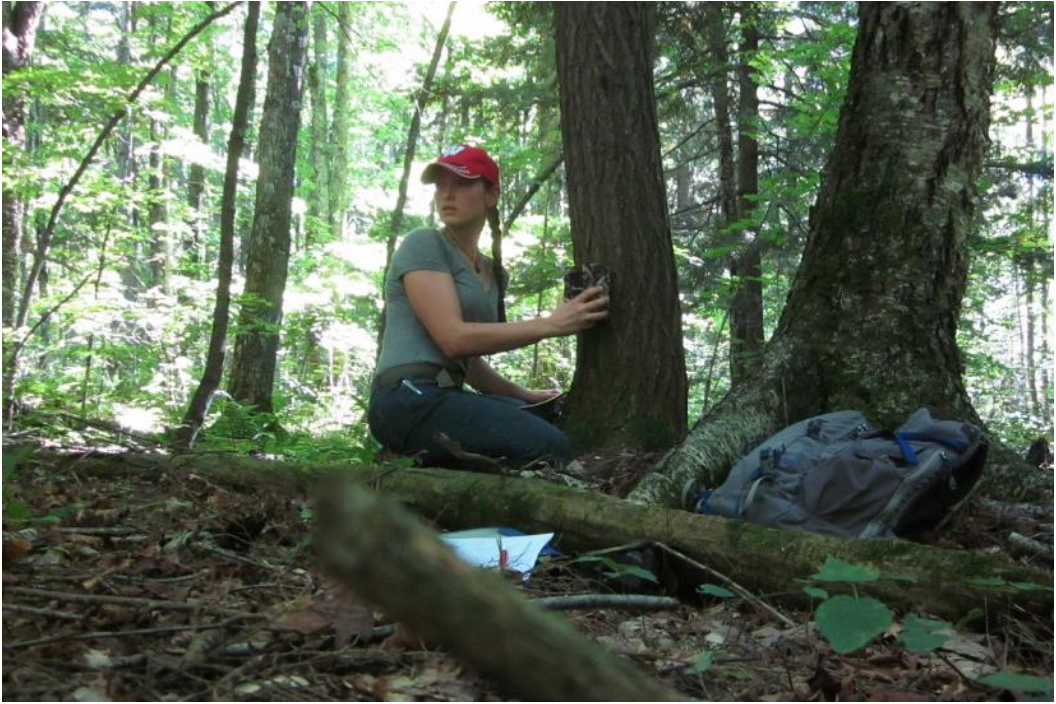

*Photo 11. Field staff placing a camera*

4) *Testing the camera*

To ensure our target location and camera placement is optimal, we took a photo on a handheld camera. We placed the camera right in front of the lens of the motion-triggered camera, allowing us to take a photo that shows the expected view of the motion-triggered camera. The goal was for the test photo to focus on the focal point, including a targeted trail junction or downed log/tree.

It was sometimes necessary to take multiple photos and adjust the placement of the camera accordingly until it was optimally placed. Alternatively, a handheld photo viewers (such as a Cuddeback® Cuddeview, Park Falls, Wisc., USA) could be used to view photos taken by the motion-triggered camera itself.

5) *Record the data*

We marked and labeled the coordinates of each camera site with a handheld GPS unit and recorded the coordinate location of the camera on datasheets. Optionally, notes were taken on habitat, animal sign seen, any potential problems with the camera, or other relevant observations. We did not place flagging or physically mark any of the camera sites in case that might influence the behavior of animals, or increase the risk of theft or damage to the camera.

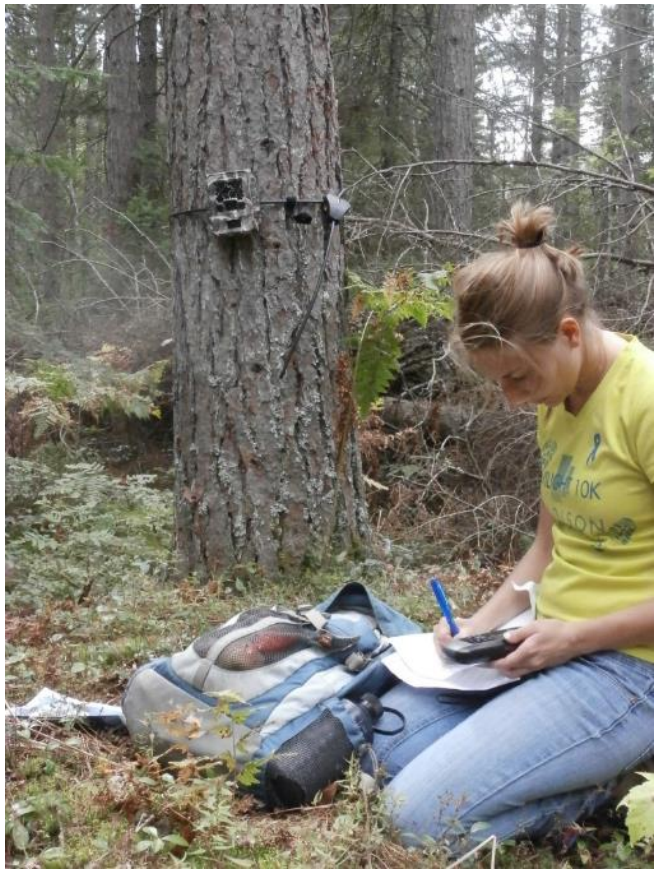

*Photo 12. Field staff recording data during a camera deployment.*

6) *Optional: Lure placement*

We speculated that placing an olfactory or visual lure could increase the probability of detection for some carnivores (Long et al. 2008). Our system for olfactory lures was to place a small amount of lure (Caven's Gusto lure, Minnesota Trapline Products, Pennock, MN) inside of a holder and hang it from a tree ~5ft high within the camera's field of view. Holders for lure can be as simple as placing the lure on a cotton ball within a Dixie cup or film canister. We also placed another small amount of lure in a crack or under bark on a log, stump, or tree within the focal area of the camera's field of view, and we placed the stick used to apply lure in the middle of the focal area.

### *Checking and Retrieving Cameras*

We left the cameras to monitor for approximately six months, and then revisited to collect SD memory cards and exchange batteries as needed. Stockton Island cameras were deployed for a second six-month interval, and then removed during the fall of 2015 to establish grids on other islands, which were also maintained for at least one full year.

Steps for retrieving data from cameras:

1) When approaching a deployed camera, walk in front of the camera to trigger a picture on the camera to provide an exact record of the date and time retrieved, and checked if the camera was still taking pictures when opened. If not, we noted this on the data sheet to help track when a camera was not functional throughout the entire monitoring period.

2) When checking a camera, adjust the setup according to your judgement if any changes had occurred (often the result of black bear investigating the cameras).

3) We then turned the camera on and checked the following data:

   Pressing 'OK' led to a screen which indicated # PICS (the number of photographs taken), %FULL (how full the memory card was) and %LITH (the available battery power remaining). We recorded #PICS & %LITH on the data sheet.

4) If the batteries were ≤90%, we replaced them.

5) We removed the SD card containing data and replaced it with a new SD card. The full SD card was stored in a paper envelope labeled with the date, the name of the camera deployment location, and the observer.

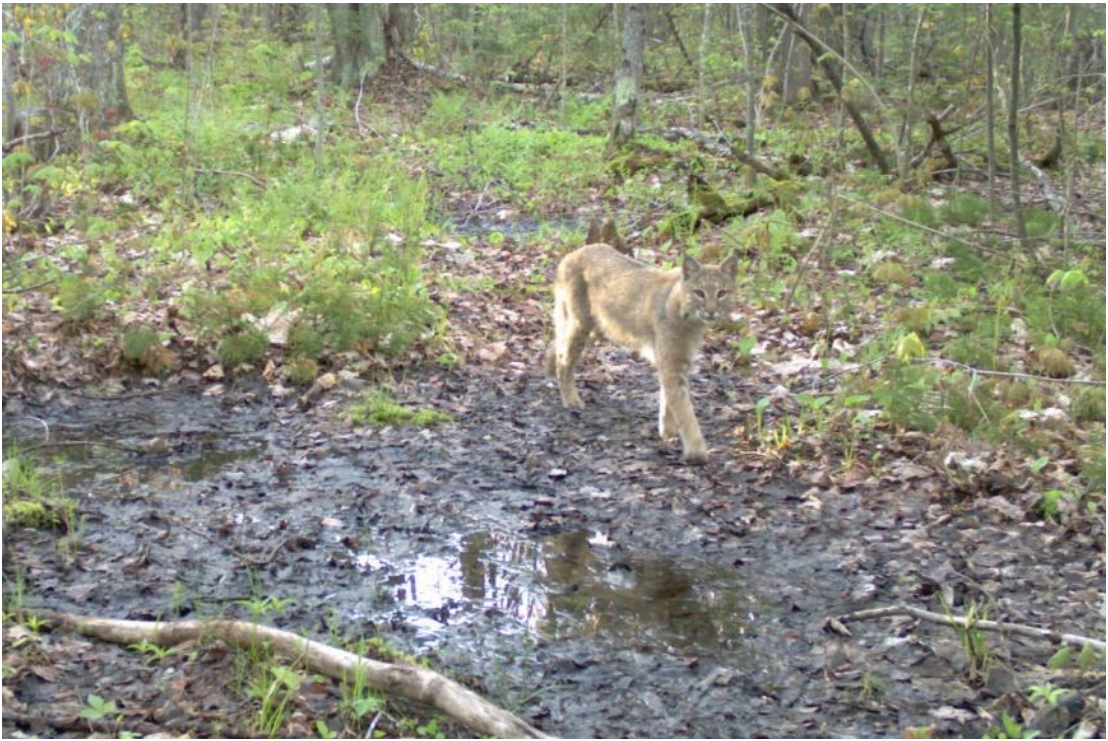

*Photo 13. A bobcat walking around an ephemeral pool on Stockton Island.*

***Camera Data Management***

All SD memory cards were labeled with the camera deployment location to avoid misidentification, and then the data were downloaded and stored in duplicate locations (which can be distributed among collaborators). We organized all photos in a database master file, backed up through cloud hosting, consisting of folders for each camera for each period of monitoring. We then tagged the individual photo files using a standardized tagging procedure (see Appendix 2).

# Appendix 2
# Protocol for Tagging Photo Data from Camera Traps

# From:
# Survey techniques, detection probabilities, and the relative abundance of the carnivore guild on the Apostle Islands (2014-2016)

# Final Report
### December 15, 2016

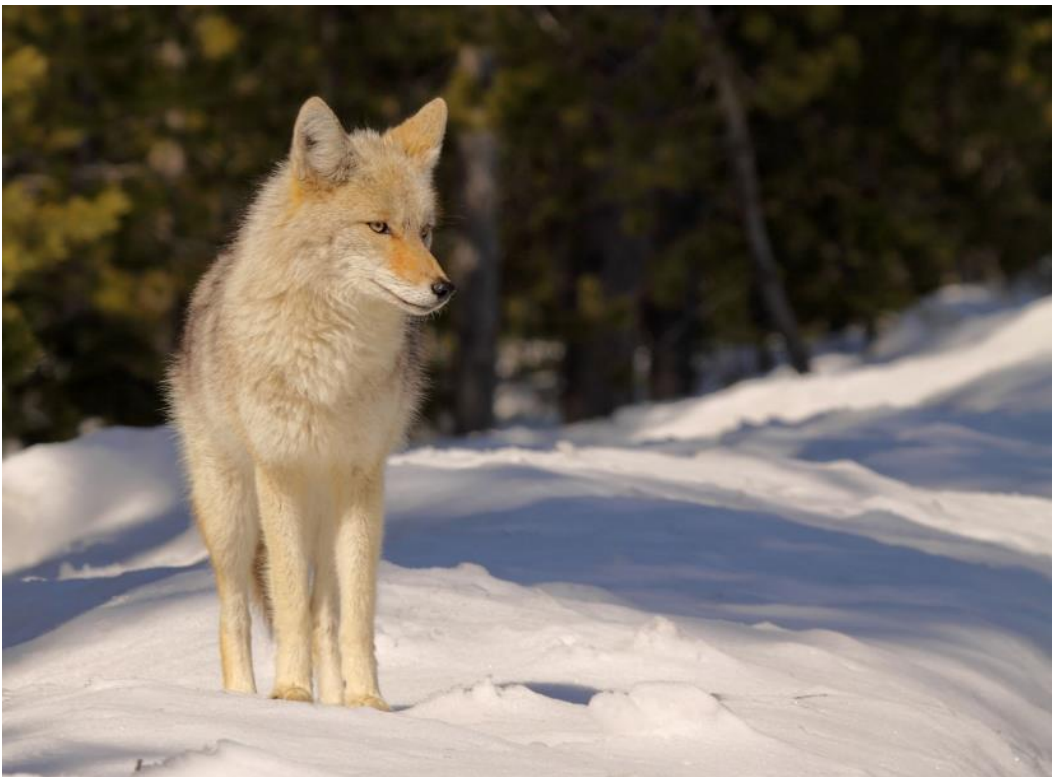

Photo by Max Allen

Suggested Citation:
Allen, M. L., B. E. Evans, M. E. Wheeler, M. A. Mueller, K. Pemble, E. R. Olson, J. Van Stappen, and T. R. Van Deelen. 2016. Survey techniques, detection probabilities, and the relative abundance of the carnivore guild on the Apostle Islands (2014-2016). Final Report to the National Park Service.

*Introduction*

This protocol uses Reconyx® proprietary software to facilitate classification of camera-trap photographs and to create a database of camera meta-data. The software is provided with the purchase of Reconyx® motion triggered trail cameras and is available online (http://www.reconyx.com/software/mapview). Use of this software does not imply an endorsement by the USDI Park Service or its collaborators.

*Setting up Reconyx Software and Loading Images*

1) Open **MapView Professional**, and click *Install* if prompted. 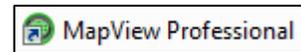

2) To confirm that image data will be linked to the correct project, navigate to the top panel of the RECONYX homepage and click **Tools - Image Folders...** and then click *Edit*. You can type the path or click the box "…" to browse.

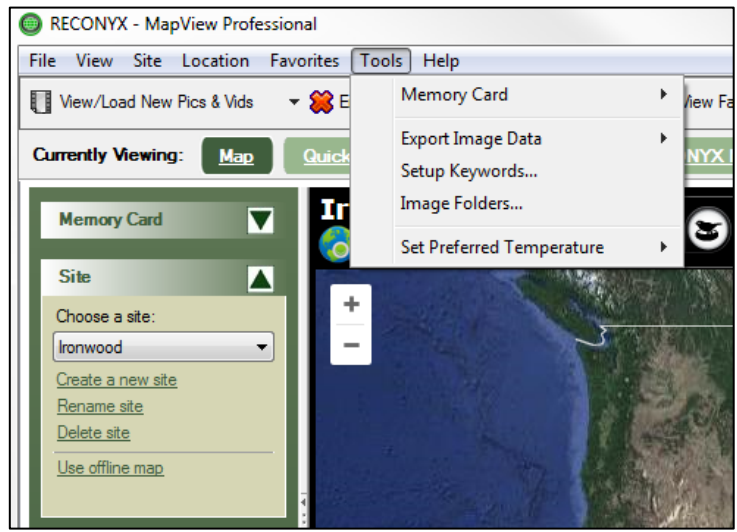

3) From the RECONYX homepage, go to the left panel and under **Choose a site** either:

   a) Select the SITE from the pull-down menu where the camera you will be tagging belongs (as an example, for the APIS project each island where cameras have been deployed is listed as a site)

   b) Or, you can create a NEW SITE and it will then be visible for the project.

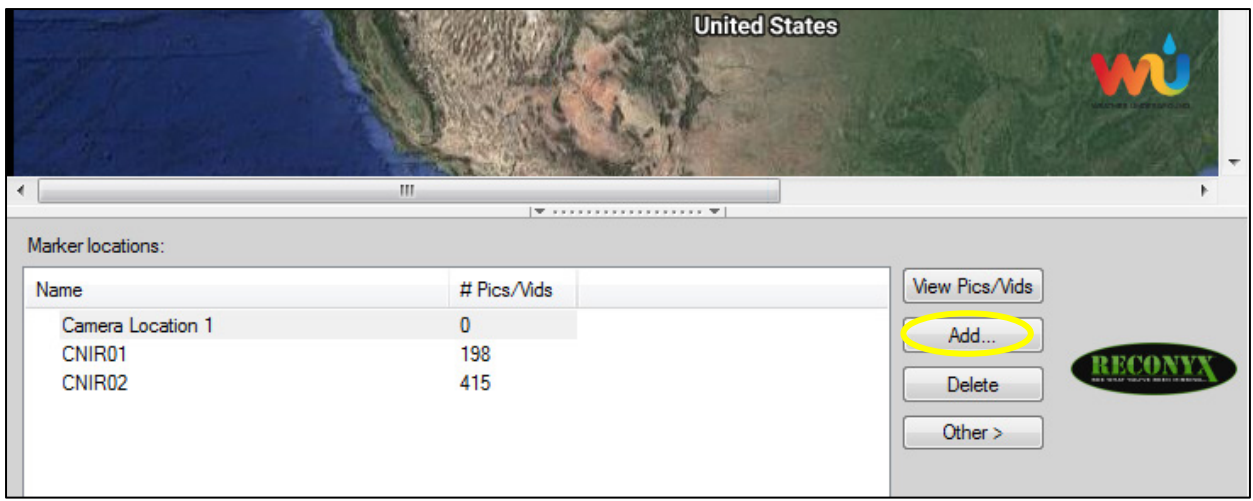

4) Back in the homepage, navigate to the lower center box. Any previously tagged cameras in that site will be visible here.

   a) To create a new camera, click the **Add** box on the right hand side. Naming conventions vary, but must be unique to each individual camera deployment location (see Appendix 1 for an example where CNBE01L and CNOA02 are two camera deployment locations).

   b) To add images, select the site and then navigate to the upper left hand console, just below **File**, and click the down arrow ▼ beside **View/Load New Images & Videos**

   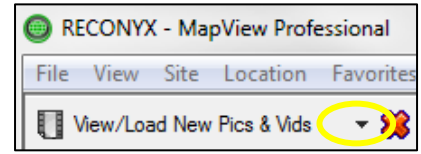

   c) Select **From Another Folder**, then navigate to your project directory, double click the camera folder, hit *OK* when prompted.

   d) In the Reconyx dialog box, select *Check All* at the top of the window, ensure the Checked/Total shows all images, and then click *OK* at the bottom. Loading may take several minutes for many images; once it is completed click *Finish* at the bottom right.

   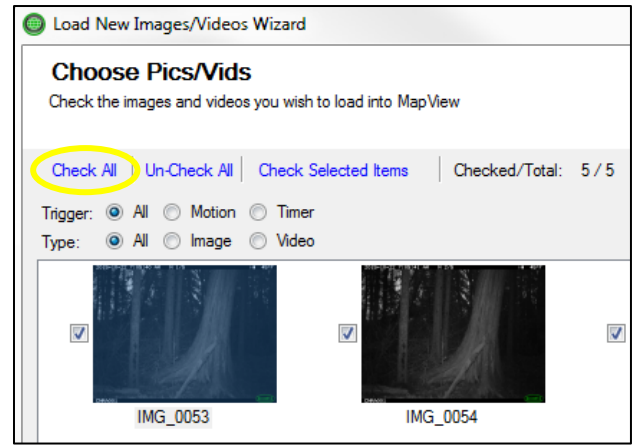

   e) If multiple folders contain images, load them sequentially into the same camera site.

Allen et al. 2016 – Apostle Island Carnivore Guild

### Creating the Keyword List and Progress Tracking Document

1) For initial setup or the first use of RECONYX software on a new computer, you must create a *Keyword List* to ensure consistent tags are applied. From the homepage, select a camera site from the lower middle window and click **View Pics/Vids** to the right

2) In the image Viewer window, go to the Image Toolbox on the right hand side and select **Setup** from the option near *Keywords* 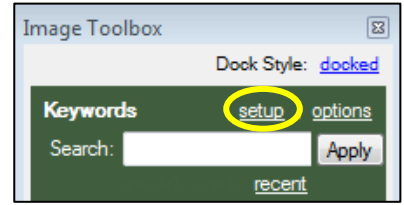

3) Click **Add** to make a new list, being careful to Name it consistently and select the correct list Type (for APIS, use *Categorical Census*, which allows a count of the number of individuals of the same species). 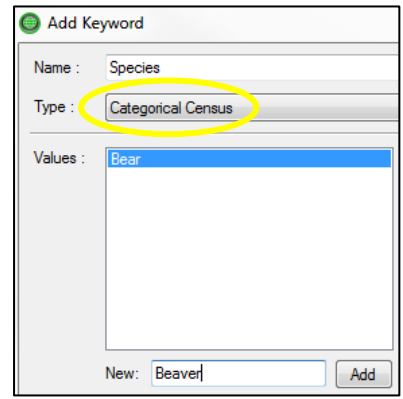

4) In the box for **New**, add each possible trigger cause for your project (these include species found in your study area, and may include other causes such as false triggers, camera setup or take down, etc.). An example list of the tags used for the APIS project is provided.

5) If you later encounter a new species, it can be added to this list. We recommend immediately updating the master list to include the new keyword, and notifying any collaborators of the addition.

6) We also recommend maintaining a Word document for tracking image tagging progress and recording information about camera site quality, uncertain animal identifications, and especially high quality images. For APIS, this document includes columns for a) Camera ID b) Image tagger initials and the dates started / completed c) List of species detected and d) Notes / Uncertain / Good. In this final column, note any days the camera was blocked by snow or vegetation, highlight any animal tags that need to be reviewed by an expert, and list file names for any particularly good images. This document can also be used to assign different individuals or teams specific camera sites to tag in order to avoid redundancy.

*Image Tagging Protocol*

1) From the RECONYX homepage, add all images to the camera site you wish to tag and then select it in the lower middle window and click **View Pics/Vids** to the right

2) In the image viewing window, individual pictures are in the column on the far left. Each image is named with the date, time, and indicator "T" for Timelapse or "M #_#" for Motion trigger. 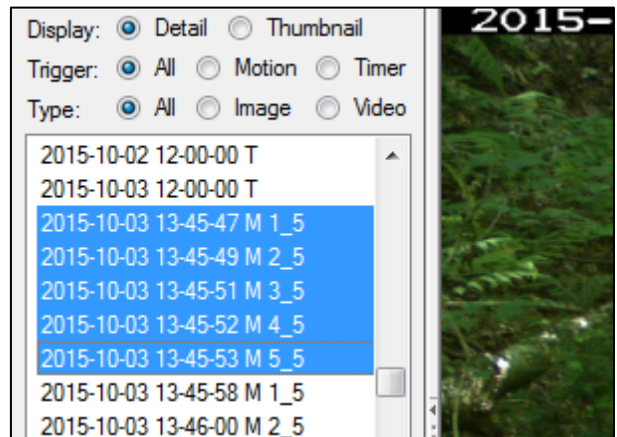

   a) Timelapse images are recorded at predetermined time each day and may or may not be present depending on the project goals and camera settings.

   b) For Motion images, the numbers indicate the order in which each image was recorded if the camera was set for a RapidFire burst. For the APIS project, bursts of five were used, so images M 1_5, M 2_5… M 5_5 result from each trigger event. Other projects may use longer bursts, or only single images per trigger event

c) All Timelapse images must be viewed and tagged either as Timelapse or as a species tag if an animal is visible. You can sort among these by selecting "All", "Timer" or "Motion" at the upper left to either tag Timelapse separately or simultaneously with Motion triggers.

3) Motion images resulting from RapidFire burst are tagged as a group.

   a) To view a series of images, click on the first image, and then use the keyboard down arrows to scroll through, examining each image for the cause of the trigger. Because several bursts of images may have the same cause, continue down the list until the cause changes.
   Example: if you have looked through 15 images featuring a squirrel and then a new set features a deer, stop on that first image of the new set.

   b) When the cause of the trigger changes, you have reached the end of that group. Hold SHIFT+UP to select back to the first image, which will be shown by blue highlighting. If the final image is not also selected, hold SHIFT+DOWN until the entire group ends is selected (for APIS, this will always mean the last image is M 5_5).

   c) DO NOT TAG INDIVIDUAL MOTION IMAGES. If RapidFire was used, always tag every image in the group consistently.
   Example: if the first two images appear blank, but a deer is visible in the final three, tag all M 1_5 to M 5_5 as Deer).

4) To apply an image tag after identifying the cause of the trigger and selecting the group, click the blue underlined **apply keyword** on the right. In the dialog box that opens, check the box to the left of the correct tag and click *OK*.

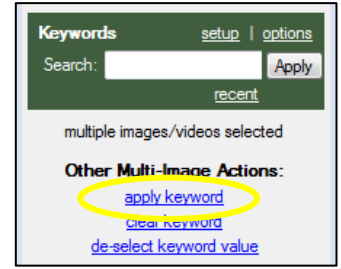

5) <u>If multiple individuals are present:</u>

   a) The default count is "1" when you click the check box beside a tag. If more than a single individual of the same species is confirmed within that burst of three images, manually apply the correct count.

   b) ONLY count the number of individuals that can be confirmed within a single burst of three images (though there may be single images where not all individuals are visible). Do not assume that the maximum visible from a previous or following trigger are present.

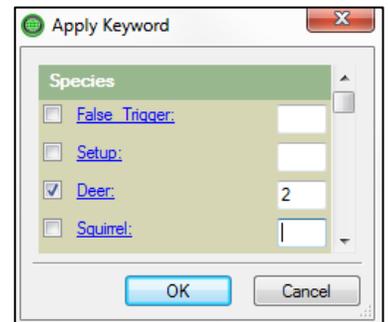

6) <u>If multiple species are present:</u>

   a) ONLY APPLY A SINGLE SPECIES TAG (do not check more than one box).

   b) Apply the tag for the larger species.

   c) Add the other species name in the Narrative box.

   d) Include a detailed account in the Word document, with the names of all images.

   Example: "For the images 2016-06-08 14:55:00 to 2016-06-08 15:03:00, a deer was present browsing in the focal area but several songbirds are also visible in the trees"

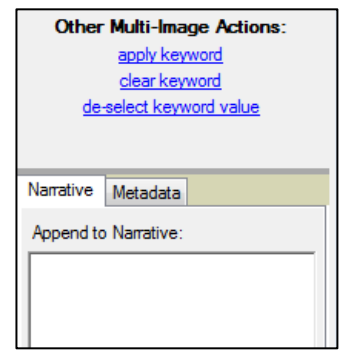

7) <u>If the trigger is cause by people:</u>

   a) For images of the deployment team setting up the camera, checking it or taking it down, apply the tag Setup. A count of individuals is not necessary here. For other human activity at the camera site, apply the tag People and record count data

   Note: You may encounter people walking dogs. For images with both dogs and humans, tag as People, for bursts with only the dog tag as Domestic_dog. Apply a Narrative comment and a note in the tracking document.

8) <u>If there is no clear cause:</u>

   a) If there are no animals present in any of the images from the event, tag as False_trigger

   b) If there is potentially a blurry animal but not enough information to identify, tag as False_trigger and make a note in the Word document with the image name and that it was unidentifiable

   c) If there is an animal but you are uncertain in your ID, apply the species tag that you believe it is and make a note in the Word document, clearly indicating that the images need to be DOUBLE-CHECKED during a quality control meeting after tagging. You can also comment in the Narrative box, but we recommend having a protocol in place to remove this comment after the images have been resolved.

9) <u>If the image is compromised:</u>

   a) Multiple things can interfere with the quality of the images (e.g., vegetation blocking the view, snow accumulation, change in the camera view point, moisture condensation on lenses, malfunctioning IR illuminator). For these, note in the Word document and include the date range, if the situation worsens or when it improves.

*Exporting Images*

When all images are tagged, before closing the window go up to **Image** and select **Export Image Data**

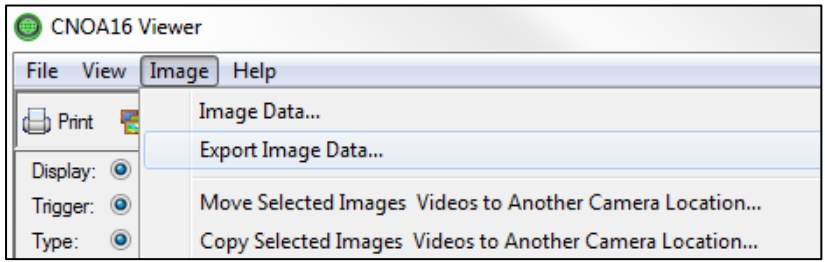 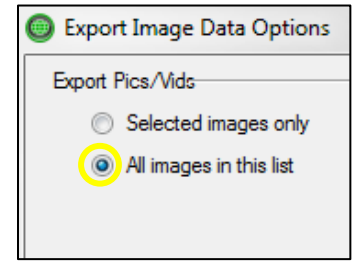

1) Switch the default to **All images in this list**

2) Save as a CSV file into the correct folder, and we suggest naming with the same conventions for making the camera site in the Reconyx software, with an additional indicator for which deployment was tagged if cameras are revisited multiple times

3) Do not close the program while it is exporting, and when prompted select **Open File** to review the exported data

    a) Check that the correct number of rows are present

    b) Go to the last column and scroll through to ensure that every row received a SPECIES tag

    c) If you had any uncertain tags, highlight these rows in the CSV file

4) Review the Word document, add any comments, and save it with the current date

5) We recommend going through the image folder and copying any uncertain images, or images of especially good quality, and save those into a special folder. To assist finding good example images later, rename these images with the species, anything of note, and the camera name. Example: "Bear with three cubs CNHE02".

**A suggested *Species List* for the APIS and surrounding area:**

| | |
|---|---|
| Bear | Fisher |
| Beaver | Fox_gray |
| Bird_corvid | Fox_red |
| Bird_goose | Fox_unknown |
| Bird_raptor | Marten |
| Bird_sandhillcrane | Otter |
| Bird_songbird | People |
| Bird_vulture | Raccoon |
| Bird_waterfowl | Rodent |
| Bird_woodcock | Setup |
| Bobcat | Snowshoe_hare |
| Cottontail | Squirrel |
| Coyote | Timelapse |
| Deer | Unknown |
| Domestic_dog | Weasel |
| False_trigger | Wolf |

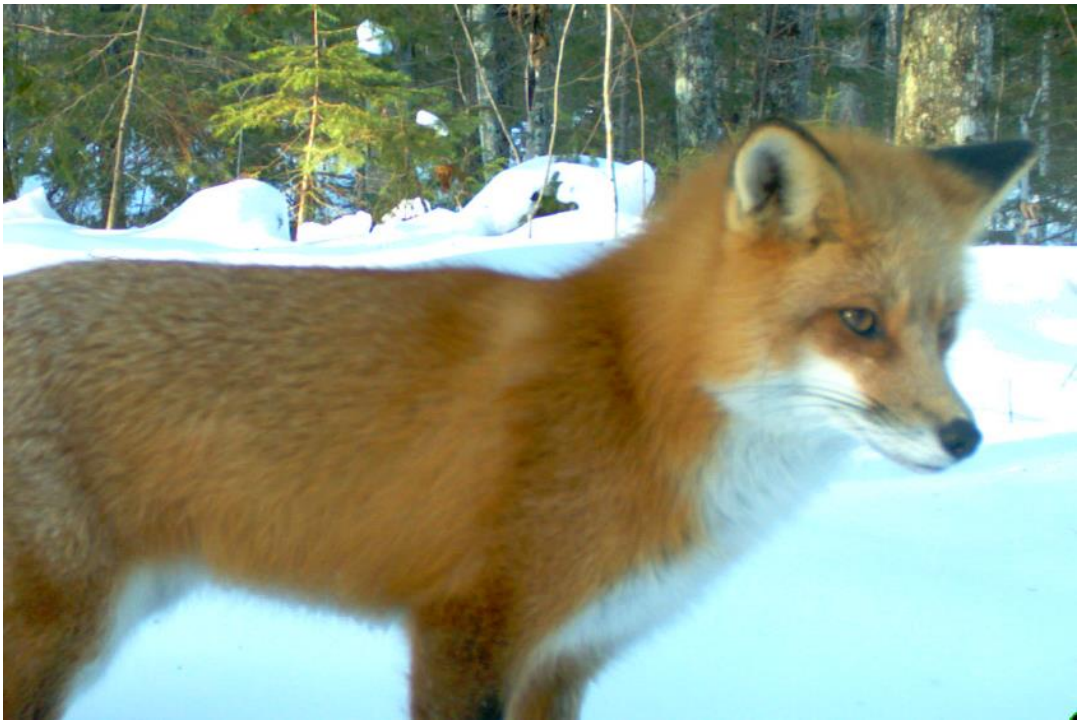

*Photo 14. A red fox pausing close to a winter camera.*